\documentclass{article}
\usepackage{amssymb}
\usepackage{amsmath}
\usepackage{amstext}

\newcommand{\beq}{\begin{equation}}
\newcommand{\eeq}{\end{equation}}
\newcommand{\beqa}{\begin{eqnarray}}
\newcommand{\eeqa}{\end{eqnarray}}
\newcommand{\nn}{\nonumber \\}
\def \podr {&& \hspace{-15pt}}

\newcounter{mtheorem}
\newcounter{mremark}
\newcounter{mdefinition}
\newcounter{mexample}
\newcounter{mexercise}

\def \setcntrs {\setcounter{equation}{0}\setcounter{mtheorem}{0}\setcounter{mremark}{0}\setcounter{mdefinition}{0}\setcounter{mexample}{0}\setcounter{mexercise}{0}}

\newcommand{\msection}[2]{%
 \refstepcounter{section}
 \section*{\large \arabic{section}. #1}\label{#2}%
 \addcontentsline{toc}{section}{\arabic{section}. #1}
 \setcntrs}
\newcommand{\msubsection}[2]{%
 \refstepcounter{subsection}
 \subsection*{\bf\normalsize \arabic{section}.\arabic{subsection}. #1}\label{#2}%
 \addcontentsline{toc}{subsection}{\arabic{section}.\arabic{subsection}. #1}}
\newcommand{\apsection}[2]{%
 \renewcommand{\thesection}{\Alph{section}}%
 \refstepcounter{section}
 \section*{\large Appendix \Alph{section}. #1}\label{#2}%
 \addcontentsline{toc}{section}{Appendix \Alph{section}. #1}
 \setcntrs}

\renewcommand{\thesection}{\arabic{section}}

\newcommand{\medskp}{

  \medskip

  }

\newenvironment{mproposition}[1][\bf Proposition \thesection.\arabic{mtheorem}]{%
	\refstepcounter{mtheorem}\medskp\noindent\textbf{#1.}\it ${}$ ${}$}{\medskp}
\newenvironment{mobservation}[1][\bf Observation \thesection.\arabic{mtheorem}]{%
	\refstepcounter{mtheorem}\medskp\noindent\textbf{#1.}\it ${}$ ${}$}{\medskp}

\newenvironment{mremark}[1][\it Remark \thesection.\arabic{mremark}]{%
	\refstepcounter{mremark}\medskp\noindent\textbf{#1.} }{\medskp}

\newcounter{tmpc}

\newenvironment{plist}{%
\setcounter{tmpc}{0}
\begin{list}{{\rm (\alph{tmpc})}}{\usecounter{tmpc}
\setlength{\leftmargin}{16pt}
\setlength{\rightmargin}{0cm}
\setlength{\itemsep}{2.5pt}
\setlength{\topsep}{5pt}
\setlength{\labelsep}{2pt}
\setlength{\labelwidth}{13pt}
\setlength{\listparindent}{18pt}}
}{\end{list}}

\newcommand{\mbf}[1]{\ensuremath{\mathchoice
                    {\mbox{\boldmath$\displaystyle\mathbf{\mathit{#1}}$}}
                    {\mbox{\boldmath$\textstyle\mathbf{\mathit{#1}}$}}
                    {\mbox{\boldmath$\scriptstyle\mathbf{\mathit{#1}}$}}
                    {\mbox{\boldmath$\scriptscriptstyle\mathbf{\mathit{#1}}$}}}}

\DeclareSymbolFont{stmry}{U}{stmry}{m}{n}
\DeclareSymbolFont{ltrsa}{OMS}{cmsy}{m}{n}
\DeclareSymbolFont{ltrs}{OT1}{pzc}{m}{it}
\DeclareMathDelimiter\llbracket{\mathopen}{stmry}{"4A}
					  {stmry}{"71}
\DeclareMathDelimiter\rrbracket{\mathclose}{stmry}{"4B}
					   {stmry}{"79}

\newcommand{\dbrackets}[1]{\left\llbracket{#1}\right\rrbracket}
\DeclareMathSymbol{\cnfalg}{\mathord}{ltrs}{"63}
\DeclareMathSymbol{\confgr}{\mathord}{ltrsa}{"43}
\DeclareMathSymbol{\compgr}{\mathord}{ltrsa}{"4B}
\def \confalg {\text{\large $\cnfalg$}}
\def \cconfalg {\confalg_{\text{\tiny $\C$}}}

\def \R {{\mathbb R}}
\def \C {{\mathbb C}}
\def \Z {{\mathbb Z}}
\def \N {{\mathbb N}}
\def \Ss {\mathcal{S}}
\def \Sr {{\mathbb S}}
\def \M {\overline{M}}
\def \VA {{\mathcal V}}
\def \spr {\cdot}
\def \SCT {C}
\def \ID {{\mathbb I}}
\def \di {\partial}
\def \dirz {z \hspace{-5pt} \text{\small $/$}}
\def \dirw {w \hspace{-5pt} \text{\small $/$}}

\def \dirv {v \hspace{-5pt} \text{\small $/$}}
\def \dira {a \hspace{-5pt} \text{\small $/$}}
\def \dirb {b \hspace{-4pt} \text{\small $/$}}
\def \dirc {c \hspace{-4pt} \text{\small $/$}}
\def \dird {d \hspace{-5pt} \text{\small $/$}}
\def \dirdi {\di \hspace{-5.5pt} /}

\def \CH {\widetilde{H}}
\def \sphi {{\phi\raisebox{7.5pt}{\hspace{-4pt}\scriptsize $*$}}}
\def \Lagr {{\mathcal L}}
\def \TRACE {{\rm tr}}
\def \Spin {\mathit{Spin}}
\def \DIM {\mathit{dim}}
\def \MOD {\mathit{mod}}

\newcommand{\cnj}[1]{{#1}^{\hspace{1pt} +}}

\newcommand{\txfrac}[2]{\frac{\raisebox{1pt}{$#1$}}{\raisebox{-3pt}{$#2$}}}
\newcommand{\nfrc}[2]{%
\text{\raisebox{1pt}{\(#1\)}\hspace{0pt}$/$\hspace{-1.5pt}\raisebox{-4pt}{\(#2\)}}}

\def \lvac {\left\langle 0 \! \left| \right. \right.  \!}
\def \rvac {\!\! \left. \left. \right| \! 0 \right\rangle}
\def \La {\left\langle \!\!{\,}^{\mathop{}\limits_{}}_{\mathop{}\limits^{}}\right.}
\def \Ra {\left. \!\!{\,}^{\mathop{}\limits_{}}_{\mathop{}\limits^{}}\right\rangle}
\def \Vl {\left. \!\!{\,}^{\mathop{}\limits_{}}_{\mathop{}\limits^{}} \right|}

\def \RA {\left.\raisebox{10pt}{\hspace{-3pt}}\right\rangle}
\def \VL {\hspace{2pt}\left.\raisebox{10pt}{\hspace{-3pt}}\right|}
\newcommand{\vrestr}[2]{\!\left.\raisebox{#1}{$\,$}\!\right|_{\,
\raisebox{1pt}{\small \(#2\)}}}
\newcommand{\gvspc}[1]{\raisebox{#1}{$\,$}$\!$}
\newcommand{\mgvspc}[1]{\raisebox{#1}{$\,$}\!}

\title{Conformal invariance and rationality in an even dimensional quantum field theory}
\author{N.M. Nikolov$^{\hspace{2pt}a}$ \ and \
I.T. Todorov$^{\hspace{2pt}a,\hspace{2pt}b,\hspace{2pt}c}$}
\date{
$^{a}$\hspace{2pt}\parbox[t]{228pt}{\small
Institute for Nuclear Research and Nuclear Energy \\
Tsarigradsko Chaussee 72, BG-1784 Sofia, Bulgaria} \\
\vspace{7pt}
$^{b}$\hspace{2pt}\parbox[t]{228pt}{\small
Institut des Hautes Etudes Scientifiques, Le Bois-Marie \\
35 route de Chartres, F-91440 Bures-sur-Yvette, France} \\
\vspace{7pt}
$^{c}$\hspace{2pt}\parbox[t]{228pt}{\small
Institut f\"ur Theoretische Physik, Universit\"at G\"ottingen \\
Tammannstrasse 1, D-37077 G\"ottingen, Germany} \\
\vspace{15pt}
{\small \today}
}

\begin{document}

\maketitle

\thispagestyle{empty}

\begin{abstract}
Invariance under finite conformal transformations in Minkowski space and
the Wightman axioms imply strong locality (Huygens principle) and rationality
of correlation functions, thus providing an extension of the concept of vertex
algebra to higher dimensions. Gibbs (finite temperature) expectation values appear
as elliptic functions in the conformal time. We survey and further pursue our
program of constructing a globally conformal invariant model of a hermitean
scalar field $\Lagr$ of scale dimension four in Minkowski space-time which can
be interpreted as the Lagrangian density of a gauge field theory.
\end{abstract}

\vfill

\indent e-mail addresses: mitov@inrne.bas.bg, todorov@inrne.bas.bg

\newpage

\tableofcontents

\newpage

\msection{Introduction}{sec1}

The present paper provides an updated survey of our work
\cite{NT01} \cite{NST02} \cite{NST03} \cite{N03} aimed at constructing a
non-trivial {\it globally conformal invariant} (GCI) 4-dimen\-sional {\it quantum
field theory} (QFT) model. Our attempt to build such a model is based on the
following results of \cite{NT01}.

Invariance under finite conformal transformations in Minkowski space-time $M$ and
local commutativity imply the {\it Huygens principle}: the commutator of two local
Bose fields vanishes for non-isotropic separations. The Huygens principle and
energy positivity yield rationality of correlation functions (Theorem 3.1 of
\cite{NT01}). These results allow to extend any GCI QFT to compactified
Minkowski space $\M$, which admits the following convenient complex variable
realization (\cite{T86} \cite{NST02} \cite{NST03} \cite{N03} \cite{NT03}):
\begin{eqnarray}
\label{eq1.1}
\hspace{-20pt}
\M = \left\{\raisebox{10pt}{\hspace{-4pt}}\right.
z \in \C^4 :
\podr
z_{\mu} = e^{2\pi i\zeta} u_{\mu} \ \text{for} \
\mu = 1 , \ldots , D\, , \ \zeta \in
\R \, ,  \nonumber \\
\podr u \in \Sr^3 \, ( \hspace{1pt} \equiv \!
\left\{\raisebox{9pt}{\hspace{-3pt}}\right.
u \in \R^4 : u^2 = {\mbf{u}}^2 + u_D^2 = 1
\left.\raisebox{9pt}{\hspace{-4pt}}\right\})
\left.\raisebox{10pt}{\hspace{-4pt}}\right\}
\cong \frac{\Sr^1 \! \times \Sr^{D-1}}{\Z_2} \, .
\end{eqnarray}
The variable $\zeta$ plays the role of {\it conformal time}.
Fields $\phi (z)$ are expressed as formal power series of the form
\begin{equation}
\label{eq1.2}
\phi \left( z \right) \, = \, \sum_{n \, \in \, {\Z}} \, \sum_{m \, \geqslant \, 0}
\left( z^{\, 2} \right)^n \, \phi_{\{n , m \}} \left( z \right)
\, , \quad
z^{\, 2} \, := \, \sum_{\mu \, = \, 1}^D z_{\mu}^2
\, ,
\end{equation}
$\phi_{\left\{ n,m \right\}} \left( z \right)$ being
(in general, multicomponent) operator valued polynomial in
$z$ which is {\it homogeneous and harmonic},
\begin{equation}
\label{n1.3}
\phi_{\{n , m \}} \left( \lambda z \right) \, = \,
\lambda^m \, \phi_{\{n , m \}} \left( z \right)
\, , \quad
\Delta \, \phi_{\{n , m \}} \left( z \right) \, = \, 0
\, , \quad
\Delta \, = \, \sum_{\mu=1}^D
\frac{\partial^2}{\partial z_{\mu}^2} \, .
\end{equation}
This is unambiguous \cite{N03}, because, as it is well known, every homogeneous
polynomial $p(z)$ of degree $m$ has a {\it unique decomposition}
\(p(z) = h(z) + z^{\, 2} q(z)\) where $h$ is harmonic (of degree $m$) and $q$
is homogeneous of degree $m-2$.

The resulting ({\it analytic}) {\it $z$-picture} provides a higher dimensional
generalization \cite{N03} of {\it chiral vertex algebras} (which have been an
outgrow of physicists' work on conformal field theory and dual resonance
models, formalized by R.E. Borcherds \cite{B86} and since subject of numerous
studies, including several books~--~see, {\it e.g.} \cite{K98} \cite{FB-Z} and
references therein).

The coordinates~(\ref{eq1.1}) are obtained
under the complex conformal transformation (with singularities)
\begin{eqnarray}
\label{eq1.5}
&& g_c : M_{\C} (\ni \xi) \to E_{\C} (\ni z)
\, , \quad
{\mbf{z}} =
\frac{\mbox{\boldmath$\xi$}}{\omega_{\xi}}
\, , \quad
z_D = \frac{1-\xi^2}{\omega_{\xi}}
\, , \nonumber \\
&& \xi^2 = {\mbox{\boldmath$\xi$}}^2 - (\xi^0)^2
\, , \quad
\omega_{\xi} = \frac{1}{2} (1 + \xi^2) - i\xi^0
\, , \quad
z^{\, 2} = \frac{1+\xi^2 + 2i\xi^0}{1+\xi^2 -2i\xi^0} \, .
\quad
\end{eqnarray}
which is regular on the \textit{Minkowski forward tube domain}
$$
{\mathcal T}_+ = \left\{ \xi = x + iy : x,y \in M , y^0 > \vert {\mbf{y}}
\vert := \sqrt{y_1^2 + \ldots + y_{D-1}^2} \right\}
$$
and maps it onto an open subset
\begin{equation}
\label{eq1.4}
T_+ =
\left\{\raisebox{14pt}{\hspace{-3pt}}\right.
z \in {\C}^D : \left\vert z^{\, 2} \right\vert < 1 \, , \ \vert z \vert^2
\left(\raisebox{10pt}{\hspace{-2pt}}\right.
:= \sum_{\mu = 1}^D \vert z^{\mu} \vert^2
\left.\raisebox{10pt}{\hspace{-2pt}}\right) < \frac{1 +
\left\vert z^{\, 2}
\right\vert^2}{2}
\left.\raisebox{14pt}{\hspace{-3pt}}\right\}
\end{equation}
of $\C^{D}$ with compact closure.
(Note that $g_c$ maps the real Minkowski space $M$ onto an open dense subset of
$\M$ (\ref{eq1.1}) such that \({\mbf{z}}^2 + (z_4 + 1)^2 = 2 e^{2\pi i
\zeta} (\cos 2 \pi \zeta + u_4) \ne 0\).)
Recall that
Minkowski space {\it spectral conditions}
(including {\it energy positivity}) imply analyticity of the
(Minkowski) vector valued function $\phi \left( \xi \right) \rvac$ for
\(\zeta \in {\mathcal T}_+\),
where $\rvac$ is the conformally invariant vacuum vector,
and leads to the {\it $z$-picture analyticity} of $\phi \left( z \right) \rvac$
for \(z \in T_+\).
It follows, in particular, that no negative powers of $z^{\, 2}$ appear in
$\phi (z) \rvac$:
\begin{equation}
\label{eq1.6} \phi_{\{n , m \}} (z) \rvac = 0
\quad \text{for} \quad n \, < \, 0 \,
\end{equation}

In the GCI~QFT the natural choice of the conformal group $\confgr$ is
the \textit{$D$-dimensional (connected) spinorial conformal group}
\(\Spin \left( D,2 \right)\) ($\equiv$ $\confgr$).
The coordinates~(\ref{eq1.1}) correspond to a special
\textit{complex} basis in the (complex) conformal Lie algebra
$\cconfalg$ including:
\begin{plist}
\item[1)]
$T_{\mu}$ for \(\mu = 1,\, \dots ,\, D\)~--~the generators of the translations
\(z \mapsto z+\lambda \, e_{\mu}\);
\item[2)]
$\SCT_{\mu}$ for \(\mu = 1,\, \dots ,\, D\)~--~the generators of the special
conformal transformations of the $z$--coordinates;
\item[3)]
\(\Omega_{\mu\nu} \equiv -\Omega_{\nu\mu}\)
for \(1 \leqslant \mu < \nu \leqslant D\)~--~the generators of the orthogonal
transformations of the $z$--coordinates.
\item[4)]
$H$, the {\it conformal Hamiltonian}~--~the generator of the $z$--dilations
$z$ $\mapsto$ $\lambda^H \left( z \right)$ $:=$ \(\lambda \, z\).
Note that $e^{2 \pi i \hspace{1pt} t H}$ acts on the conformal time
variable $\zeta$ in~(\ref{eq1.1}) by a translation
\(\zeta \mapsto \zeta+t\).
\end{plist}
The conjugation corresponding to the real conformal Lie algebra is
an antilinear {\it automorphism} acting on these generators as:
\beq\label{cnj}
T_{\mu}^{\hspace{1pt} \text{\large $\star$}} \, := \, \SCT_{\mu}
\, , \quad
H^{\hspace{1pt} \text{\large $\star$}} \, = \, - \, H
\, , \quad
\Omega_{\mu\nu}^{\hspace{1pt} \text{\large $\star$}} \, = \, \Omega_{\mu\nu}
.
\eeq
(By contrast, the hermitean conjugation is an algebra antiautomorphism
acting with an opposite sign on the generators~--~see (\ref{equ2.13}) below.)
The generators $T_{\mu}$, $\Omega_{\mu\nu}$ and $H$ span the subalgebra
$\confalg_{\infty}$ of $\confalg$, the
{\it stabilizer of the central point at infinity} in the
$z$--chart which is isomorphic
to the complex Lie algebra of Euclidean transformations with dilations.
Its conjugate $\confalg_0$,
\beq\label{c0}
\confalg_0 \, :=  \, \mathit{Span}_{\C} \,
\left\{ \SCT_{\mu},\, \Omega_{\mu\nu},\, H \right\}_{\mu,\nu}
\, ,
\eeq
is the stabilizer of \(z=0\) in
$\cconfalg$.
The {\it real counterpart} of the intersection of
$\confalg_{\infty}$ and $\confalg_0$ is the Lie algebra of the
maximal compact subgroup \(\compgr = \nfrc{\Spin(D) \times U(1)}{\Z_2}\)
of the real conformal group $\confgr$ generated by $\Omega_{\mu\nu}$ and $iH$.
The $z$-picture domain $T_+$ is a homogeneous space of the (real) group
$\confgr$ with stabilizer of $z = 0$ equal to its maximal compact subgroup
$\compgr$, so that \(T_+ = \confgr / \compgr\).
The compactified Minkowski space $\M$ (\ref{eq1.1}),
appearing as a $D$--dimensional submanifold of the boundary of  $T_+$,
is also a homogeneous space of $\confgr$.
The conformal Hamiltonian $H$ is nothing but the generator of the
centre $U(1)$ of $\compgr$.

Free massless fields and the stress energy tensor only satisfy the Huygens
principle for even space-time dimension. We shall therefore assume in what follows
that $D$ {\it is even}.

The $D$-dimensional vertex algebra of GCI fields with rational correlation
functions corresponds to the {\it algebra of local observables} in Haag's
approach \cite{H} to QFT. Its {\it isotypical} (or {\it factorial}) {\it
representations} ({\it i.e.} multiples with a finite multiplicity of an
irreducible representation) give rise to the {\it superselection sectors} of the
theory. The intertwiners between the vacuum and other superselection sectors are
higher dimensional counterparts of primary fields (which have, as a rule,
fractional dimensions and multivalued $n$-point distributions).

We shall add to the traditional assumption that the conserved (sym\-metric,
traceless) {\it stress-energy tensor} $T_{\mu\nu} (z)$ is a local observable the
requirement that so is the scalar, gauge invariant {\it Lagrangian density}
$\Lagr (z)$ (of dimension $d=D$). Then the construction of a GCI QFT
model becomes a rather concrete program of writing down rational (conformally
invariant and ``crossing symmetric'') correlation functions and studying the
associated {\it operator product expansions} (OPE).
We are pursuing this program for the $D=4$ case (of main physical interest) in
our work \cite{NST02} \cite{NST03} reviewed and continued in \cite{NST04}
(see also secs.~\ref{sec4}--\ref{sec6n} below).

We begin, in Sect.~\ref{sec2}, with a review of properties of $z$-picture fields
(or vertex operators) which follow from GCI and Wightman axioms.
Free compact picture fields are shown to have doubly periodic
(meromorphic)~--~{\it i.e.} elliptic -- correlation functions in the conformal time
differences $\zeta_{ij}$ with periods $1$ (in our units, the radius of the universe
divided by the velocity of light) and the {\it modular parameter} $\tau$ whose
imaginary part is proportional to the inverse absolute temperature
(Sect.~\ref{ssec3.1}).
These results are expected to hold beyond the free field examples \cite{NT03}.
In (Sect.~\ref{ssec3.2}) we compute energy mean values of free fields in an
equilibrium state characterized by the Kubo-Martin-Schwinger (KMS) property.
The result is expressed as a linear combination of modular forms.
The concise survey (in Sects.~\ref{sec2} and \ref{sec3}) of the results of
\cite{N03} \cite{NT03} is followed
in Sect.~\ref{sec4} with the first step in constructing a (non free) GCI model
satisfying Wightman axioms \cite{SW}. After a brief review (in Sect.~\ref{ssec4.1})
of the
results of \cite{NT01}, reformulated in the analytic $z$-picture, we write down
(in Sect.~\ref{ssec4.2}) the general truncated GCI 4-point function
$w_4^t$ of a neutral scalar field of (integer) dimension $d$.
This is a homogeneous rational function of degree $-2d$ in the
(complex) Euclidean invariant variables
\begin{equation}
\label{eq1.7}
\rho_{ij} = z_{ij}^{\, 2} \, , \ z_{ij} = z_i - z_j \, , \ z^{\, 2} = {\mbf{z}}^2 +
z_4^2
\end{equation}
of denominator
\((\rho_{12} \rho_{13} \rho_{14} \rho_{23} \rho_{24} \rho_{34})^{d-1}\)
and numerator, a homogeneous polynomial of degree $4d-6$ (for
$d \geqslant 2$) depending (linearly) on
$\dbrackets{\txfrac{d^2}{3}}$ ({\it i.e.}
no more than $d^2 / 3$) real parameters. It is just
\(c (\rho_{13} \rho_{24} + \rho_{12} \rho_{34} + \rho_{14} \rho_{23})\)
for the simplest candidate for a
non-trivial, $d=2$, model, and involves 5 parameters for the physically
interesting case of a $d=4$ Lagrangian density $\Lagr$.
As the model of a $d=2$ scalar
field was proven in \cite{NST02} to correspond to normal products of free
(massless) scalar fields we concentrate in the rest of the paper on the $d=4$
case. We study in Sect.~\ref{sec5} OPE organized in bilocal fields of fixed twist
which provide what could be called a {\it conformal partial wave expansion} of the
4-point function (a concept, introduced in \cite{DMPPT}, see also \cite{3}, and
recently revisited in \cite{5}). The bilocal field $V_1 (z_1 , z_2)$
of (lowest) dimension $(1,1)$, which admits a Taylor expansion in
$z_{12}$ involving only twist 2 symmetric traceless tensors, is harmonic in each
argument allowing to compute (in Sect.~\ref{ssec5.2}) its (rational) 4-point
function. The
corresponding (crossing) symmetrized contribution to $w_4^t$ gives rise to a
3-parameter sub-family of the original 5-parameter family of GCI 4-point
functions.
We give a precise definition of of the symmetrization ansatz in Sect.~\ref{ssec6.1n}.

We argue in Sect.~\ref{ssec6.3n} (summarizing results of \cite{NST03}), that the
Lagrangian $\Lagr (z)$ of a gauge field theory should have vanishing odd
point functions and should not involve a $d=2$ scalar field in its OPE.
This reduces
to 3 the 5 parameters in $w_4^t$. One of the remaining parameters corresponds to
the Lagrangian ({\it i.e.}, the contracted normal square) of a free Maxwell field
$F_{\mu\nu}$ giving rise to a bilocal combination of $F$'s, $V_1^{(2)}$.
The other bilocal field, $V_1^{(1)}$, contributing to the (restricted) $w_4^t$,
has the properties of a bilinear
combination of a free Weyl spinor and its conjugate (Sect.~\ref{ssec5.3n}).
This permits the computation of higher point functions as displayed in
Sect.~\ref{ssec6.2n}.

The results and the challenging open problems are discussed in Sect.~\ref{sec7}.

\msection{Vertex algebras, strong locality, rationality (a synopsis)}{sec2}

Wightman axioms \cite{SW} and GCI can be expressed as
{\it vertex algebra properties of $z$-picture fields},
\cite{N03} \cite{NT03}, which we proceed to sum up.

\msubsection{Properties of $z$-picture fields}{ssec2.1}

1) The {\it state space} $\VA$ of the theory is a ({\it pre-Hilbert})
{\it inner product space} carrying a (reducible) unitary
{\it vacuum representation} $U(g)$ of the conformal group $\confgr$, for which:

\smallskip

\noindent
1a) the corresponding representation of the complex Lie algebra $\cconfalg$
is such that the spectrum of the $U(1)$ generator $H$
belongs to \(\left\{\raisebox{10pt}{\hspace{-2pt}}\right.
0,\, \txfrac{1}{2},\, 1,\,
\txfrac{3}{2},\, \dots
\left.\raisebox{10pt}{\hspace{-2pt}}\right\}\)
and has a finite degeneracy:
\begin{equation}
\label{eq2.1new}
\VA = \bigoplus_{\rho \, = \, 0 , \frac{1}{2} , 1 , \ldots}
\VA_{\rho} \, , \ (H-\rho) \, \VA_{\rho} = 0
\, , \quad
\DIM \, \VA_{\rho} < \infty \, ,
\end{equation}
each $\VA_{\rho}$ carrying a fully reducible representation
of $\Spin \left( D \right)$ (generated by \(\Omega_{\mu\nu}\)).
Moreover, the central element $-\ID$ of the subgroup
$\Spin \left( D \right)$
is represented on $\VA_{\rho}$ as $\left( -1 \right)^{2\rho}$.

\smallskip

\noindent
1b) The {\it lowest energy space} $\VA_0$ is 1-dimensional:
it is spanned by the (norma\-lized) {\it vacuum vector} $\rvac$,
which is invariant under the full conformal group $\confgr$.

\medskip

As a consequence (see~\cite{N03}, Sect.~7) the Lie subalgebra
$\confalg_0$~(\ref{c0})
of $\cconfalg$ has locally finite action on $\VA$, i.e.,
every \(v \in \VA\) belongs to a finite dimensional
subrepresentation of $\confalg_0$.
Moreover, the action of $\confalg_0$ is integrable to an action
of the complex Euclidean group with dilations $\pi_0 \left( g \right)$
and the function
\beq\label{def_pi}
\pi_z \left( g \right) \, := \,
\pi_0 \left( t_{g \left( z \right)}^{-1} \, g \, t_z \right)
\, \
\eeq
is rational in $z$ with values in $\mathit{End}_{\C} \, \VA$
(the space endomorphisms of $\VA$)
and satisfies the cocycle property
\beq\label{cocycle}
\pi_z \left( g_1 g_2 \right) \, = \,
\pi_{g_2 \left( z \right)} \left( g_1 \right)
\pi_z \left( g_2 \right)
\quad \text{iff} \quad
g_1 g_2 \left( z \right),\, g_2 \left( z \right) \, \in \, \C^D
\, . \
\eeq

\medskip

\noindent
2) The fields \(\phi \left( z \right)
\equiv \left\{ \phi_a \left( z \right) \right\}\)
(\(\psi \left( z \right)
\equiv \left\{ \psi_b \left( z \right) \right\}\), etc.)
are represented by infinite
power series of type~(\ref{eq1.2}) and
\beq\label{equ2.4}
\phi_{n,m} \left( z \right) \, v \, = \, 0
\eeq
for all \(m = 0,1,\dots\) if \(n > N_v \in \Z\).

\medskip

\noindent
3) {\it Strong locality}:
The fields $\phi$, $\psi$,~$\dots$ are assumed to have $\Z_2$--parities
$p_{\phi}$, $p_{\psi}$,~$\dots$ (respectively) such that
\begin{equation}
\label{equ2.5}
\rho_{12}^N \,
\left\{
\phi_a \left( z_1 \right) \psi_b \left( z_2 \right)
- (-1)^{p_{\phi} p_{\psi}} \,
\psi_b \left( z_2 \right) \phi_a \left( z_1 \right)
\raisebox{9pt}{\hspace{-2pt}}
\right\} \, = \, 0 \,
\quad (\rho_{12} \, := \, z_{12}^{\, 2})
\end{equation}
for sufficiently large $N$.

\medskip

The assumption that the field algebra is $\Z_2$ graded, which underlies 3),
excludes the so called ``Klein transformations''
(whose role is discussed e.~g. in~\cite{SW}).

Strong locality implies an analogue of the
\textit{Reeh--Schlieder theorem}, the separating property of the
vacuum, namely

\begin{mproposition}
\begin{plist}
\item[{\rm (}a{\rm )}] {\rm (\cite{NT03}, Proposition 3.2~(a).)}
The series $\phi_a \left( z \right) \rvac$ does not contain negative
powers of $z^{\, 2}$.
\item[{\rm (}b{\rm )}] {\rm (\cite{N03}, Theorem~3.1.)}
Every local field component $\phi_a \left( z \right)$ is
uniquely determined
by the vector \(v_a = \phi_a \left( 0 \right) \rvac\).
\item[{\rm (}c{\rm )}] {\rm (\cite{N03}, Theorem~4.1 and Proposition~3.2.)}
For every vector \(v \in \VA\) there exists unique
local filed $Y \left( v,\, z \right)$ such that
\(Y \left( v, 0 \right) \rvac = v\). Moreover, we have
\begin{equation}
\label{eq2.3new}
Y(v,z) \rvac \, = \, e^{z \spr T} \, v
\, , \quad
z \spr T \, = \, z^1 \, T_1 + \dots + z^D \, T_D
\, . \
\end{equation}
\end{plist}
\end{mproposition}

The part (\textit{c}) of the above proposition is the higher dimensional
analogue of the {\it state field correspondence}.

\medskip

\noindent
6) \textit{Covariance.}
\beqa
\label{equ2.7}
\left[ \hspace{1pt} T_{\mu} \, , \,
Y \left( v,\, z \right) \, \right]
\, = \podr
\frac{\di}{\di z^{\mu}} \,
\, Y \left( v,\, z \right)
\, , \quad
\\ \label{equ2.8}
\left[ \hspace{1pt} H \, , \,
Y \left( v,\, z \right) \, \right]
\, = \podr
z \spr
\frac{\di}{\di z} \,
\, Y \left( v,\, z \right) \, + \,
Y \left( H\hspace{1pt} v,\, z \right)
\, , \quad
\\ \label{equ2.9}
\left[ \hspace{1pt} \Omega_{\mu\nu} \, , \,
Y \left( v,\, z \right) \, \right]
\, = \podr
z^{\mu} \,
\frac{\di}{\di z^{\nu}} \,
Y \left( v,\, z \right) \, - \,
z^{\nu} \,
\frac{\di}{\di z^{\mu}} \,
Y \left( v,\, z \right) \, + \,
Y \left( \Omega_{\mu\nu}\hspace{1pt} v,\, z \right)
\, , \quad \mgvspc{12pt}
\\ \label{equ2.10}
\left[ \hspace{1pt} \SCT_{\mu} \, , \,
Y \left( v,\, z \right) \, \right]
\, = \podr
\left(\raisebox{10pt}{\hspace{-2pt}}\right.
-z^{\, 2} \,
\frac{\di}{\di z^{\mu}} \,
+ 2 \, z^{\mu} \, z \spr
\frac{\di}{\di z}
\left.\raisebox{10pt}{\hspace{-2pt}}\right)
Y \left( v,\, z \right) \, + \,
2 \, z^{\mu} \, Y \left( H\hspace{1pt} v,\, z \right) \, + \,
\mgvspc{12pt}
\nn \podr
+ \, 2 \mathop{\sum}\limits_{\nu \, = \, 1}^D
z^{\nu} \, Y \left( \Omega_{\nu\mu} \hspace{1pt} v,\, z \right) \, + \,
Y \left( \SCT_{\mu}\hspace{1pt} v,\, z \right)
\, . \qquad
\eeqa

\medskip

Vectors \(v \in \VA\) for which \(\SCT_{\mu} v =0 \) (\(\mu = 1,\dots,D\))
are called \textbf{quasiprimary}.
Their linear span decomposes into irreducible representations
of the maximal compact subgroup $\compgr$, each of them characterized by
weights \(\left(\raisebox{9pt}{\hspace{-2pt}}\right.
d;\, j_1,\dots,j_{\frac{D}{2}} \left.\raisebox{9pt}{\hspace{-2pt}}\right)\).
We assume that our basic fields $\phi_a$, $\psi_b$,~$\dots$, correspond to
such quasiprimary vectors so that the transformation
laws~(\ref{equ2.7})--(\ref{equ2.10}) give rise to $\compgr$--induced
representations of the conformal group~$\confgr$.

If \(v \in \VA\) is an eigenvector of $H$ with eigenvalue $d_v$
then Eq.~(\ref{equ2.8}) implies that the field $Y \left( v, z \right)$
has dimension $d_v$:
\beq\label{ne2.11}
\left[ \hspace{1pt} H \, , \,
Y \left( v,\, z \right) \, \right]
\, = \,
z \spr
\frac{\di}{\di z} \,
\, Y \left( v,\, z \right) \, + \,
d_v \, Y \left( v,\, z \right)
\, . \quad
\eeq
It follows also from the correlation between the dimension and the spin
in the property~1a) and, on the other hand, the spin and statistic theorem
that the $Z_2$--parity $p_v$ of $v$ is related with its dimension
as \(p_v \equiv 2d_v \ \MOD \, 2\) and hence
\beq\label{loc}
\rho_{12}^{\mu \left( v_1,v_2 \right)}
\left\{
Y \! \left( v_1, z_1 \right) Y \! \left( v_2, z_2 \right)
- (-1)^{4 \, d_{v_1} d_{v_2}} \,
Y \! \left( v_2, z_2 \right) Y \! \left( v_1, z_1 \right)
\raisebox{10pt}{\hspace{-2pt}}
\right\} = 0
\eeq
where $\mu \left( v_1,v_2 \right)$ depends on the spin and dimensions of
$v_1$ and~$v_2$.

\medskip

\noindent
7) \textit{Conjugation.}
\beq\label{equ2.11}
\La v_1 \Vl Y \left( \cnj{v},\, z \right) v_2 \Ra \, = \,
\La Y \left(\raisebox{9pt}{\hspace{-2pt}}\right.
\pi_{z^*} \hspace{-2pt} \left( J_W \right)^{-1} v,\, z^*
\left.\raisebox{9pt}{\hspace{-2pt}}\right) v_1 \Vl v_2  \Ra
\,
\eeq
for every \(v,v_1,v_2 \in \VA\), where
\beq\label{star}
z^{\, *} \, := \, \frac{\overline{z}}{\overline{z}^{\, 2}}
\eeq
is the $z$--picture conjugation (leaving invariant the real space~(\ref{eq1.1}))
and $J_W$ is the
element of $\confgr_{\C}$ representing the so called \textit{Weyl reflection}
\beq\label{equ2.12}
J_W \left( z \right) \, := \, \frac{R_D \left( z \right)}{z^{\, 2}}
\, , \quad
R_{\mu} \left( z^1,\, \dots ,\, z^D \right) \, := \,
\left( z^1,\, \dots ,\, -z^{\mu},\, \dots ,\, z^D \right)
\, \
\eeq
($J_W^2$ being thus central element of $\confgr$).
Note that the hermitean conjugation $X^*$ of the conformal group generators
\(X \in \cconfalg\) is connected by the conjugation~(\ref{cnj})~as
\beq\label{equ2.13}
X^* \, = \, - X^{\hspace{1pt} \text{\large $\star$}}
\, . \
\eeq

\medskip

\noindent
8) {\it Borcherds' OPE relation}. The equality
\begin{equation}
\label{eq2.8new}
Y (v_1 , z_1) \, Y (v_2 , z_2)
\, \text{``} \! = \! \text{''} \,
Y (Y (v_1 , z_{12}) \, v_2 , z_2) \, ,
\end{equation}
is satisfied after applying some transformations to the formal power series
on both sides which are not defined on the corresponding series' spaces
(see~\cite{N03}, Theorem~4.3).
On the other hand, the vector valued function
$$
Y (v , z_1) \, Y (v_2 , z_2) \rvac = Y (Y (v_1 , z_{12}) \, v_2 , z_2) \rvac
$$
is analytic with respect to the Hilbert norm topology for
$\vert z_2^{\, 2} \vert < \vert z_1^{\, 2} \vert < 1$
and sufficiently small $\rho_{12}$.

\msubsection{Free field examples}{ssec2.2}

For a {\it scalar field} $\phi$ of dimension $d$ we can write
\begin{equation}
\label{eq2.9new}
\phi (z) = Y (\vert d ; 0 , \ldots , 0 \rangle , z)
\, , \quad
\text{where} \quad
\vert d ; 0 , \ldots , 0 \rangle = \phi_{\{ 0 , 0 \}} \rvac ;
\end{equation}
here $(j_1 , \ldots , j_{\frac{D}{2}})$ stands for the weight of an irreducible
$\Spin (D)$ representation, $(0 , \ldots , 0)$ labeling the trivial
(1-dimensional) one.
Its conformal invariant two point Wightman function is
\beq\label{sc_2pt}
\lvac \phi \left( z_1 \right) \phi^* \left( z_2 \right) \rvac
\, = \,
B_{\phi} \, \rho_{12}^{-d}
\, . \
\eeq
When
$d$ takes its canonical value $d_0$ for which
$\phi = \varphi (z)$ is harmonic
\begin{equation}
\label{eq2.10new}
d \, = \, d_0 \, := \, \frac{D - 2}{2}
\quad \Leftrightarrow \quad
\Delta \, \varphi (z) = 0 \, ,
\end{equation}
the expansion (\ref{eq1.2}) takes the form
\begin{equation}
\label{eq2.11new}
\varphi (z) = \sum_{\ell = 0}^{\infty} \left\{
\varphi_{-\ell - d_0} (z) +
\left( z^{\, 2} \right)^{-\ell - d_0} \, \varphi_{\ell + d_0} (z)
\raisebox{10pt}{\hspace{-2pt}}\right\}
\,
\end{equation}
where $\varphi_n (z)$ is a homogeneous harmonic polynomial of degree
$\vert n \vert - d_0$ ($\geqslant 0$) and, for a hermitean $\varphi$,
\begin{equation}
\label{eq2.12new}
(\varphi_n (\overline{z}))^* = \varphi_{-n} (z) \, .
\end{equation}
The modes $\varphi_n$ are related to $\phi_{\{n , m \}}$ of (\ref{eq1.2}) by
\begin{equation}
\label{eq2.13new}
\varphi_{-\ell - d_0} (z) = \phi_{\{ 0 , \ell \}} (z)
, \quad
\varphi_{\ell + d_0} (z) = \phi_{\{ - \ell - d_0 , \ell \}} (z)
\quad
(\, [H , \varphi_n (z)] = -n \, \varphi_n (z) \, ) .
\end{equation}

We proceed to the description of a {\it free Weyl field} for $D=4$.
Compact picture spinor fields are conveniently studied using the $2 \times 2$
matrix realization of the quaternionic algebra.
We express the imaginary quaternion units $Q_j$ in terms of the Pauli matrices
setting $Q_j = -i \, \sigma_j$ and denote by $Q_4$ the $2 \times 2$ unit matrix.
To each (complex) 4-vector $z$ we make correspond a pair of conjugate quaternions
\begin{equation}
\label{eq2.14new}
\dirz \, := \, \sum_{\mu = 1}^4 z_{\mu} \, Q_{\mu} \, = \,
z_4 + {\mbf{z}} \, {\mbf{Q}}
, \quad
\dirz^+ \, := \, \sum_{\mu = 1}^4 z_{\mu} \, Q_{\mu}^+ \, = \,
z_4 - {\mbf{z}} \, {\mbf{Q}}
\end{equation}
where
$$
{\mbf{z}} \spr {\mbf{Q}}
\, = \, \sum_{j=1}^3 z_j \, Q_j \, = \,
-i \begin{pmatrix} z_3 &z_1 - iz_2 \\ z_1 + iz_2 &-z_3 \end{pmatrix} \, .
$$
The basic anticommutation relations for quaternion 4-vectors read:
\begin{equation}
\label{eq2.15new}
\dirz \, \dirw^+ \, + \, \dirw \, \dirz^+ \, = \,
2zw \, ( \, = \, 2({\mbf{z}} \spr {\mbf{w}} \, + \, z_4 \, w_4))
\, = \, \dirz^+ \, \dirw + \dirw^+ \, \dirz \, .
\end{equation}

A {\it free Weyl field} $\psi (z)$ is a complex 2-component spinor field of
$\compgr = S(U(2) \times U(2))$ weight
$\left( \txfrac{3}{2} ; \txfrac{1}{2} , 0 \right)$, satisfying the Weyl equation:
\begin{equation}
\label{eq2.16new}
\dirdi \, \psi (z) \, := \,
Q_{\mu} \, \frac{\partial}{\partial z_{\mu}} \, \psi (z)
\, = \, 0 \
( \,= \frac{\partial}{\partial z_{\mu}} \, \psi^+ (z) \, Q_{\mu} \, ) \, .
\end{equation}
The unique conformal invariant two point Wightman function
(up to normalization) is
\beq\label{w_2pt}
\lvac \psi \left( z_1 \right) \psi^+ \left( z_2 \right) \rvac
\, = \,
\rho_{12}^{-2} \, \dirz^+_{12}
\, . \
\eeq
The
mode expansion reads:
\beqa\label{eq2.17new}
& \psi (z) \, = \, \mathop{\sum}\limits_{\ell = 0}^{\infty}
\left\{
\psi_{-\ell - \frac{3}{2}} (z) + \left( z^{\, 2} \right)^{-\ell -2} \,
\dirz^+ \, \psi_{\ell + \frac{3}{2}} (z)
\raisebox{10pt}{\hspace{-2pt}}\right\}
\, , \
& \nn &
\psi^+ (z) \, = \, \mathop{\sum}\limits_{\ell = 0}^{\infty}
\left\{
\psi_{-\ell - \frac{3}{2}}^+ (z) + \left( z^{\, 2} \right)^{-\ell -2} \,
\psi_{\ell + \frac{3}{2}}^+ (z) \, \dirz^+
\raisebox{10pt}{\hspace{-2pt}}\right\}
\, , \
&
\eeqa
where $\psi_{\rho}^{(+)} (z)$ are 2-component spinor-valued homogeneous (harmonic)
polynomials of degree $\vert \rho \vert - \txfrac{3}{2}$
($\geqslant 0$) satisfying appropriate Weyl equations,
\begin{equation}
\label{eq2.18new}
\dirdi \, \psi_{\rho} (z) \, = \, 0 \, = \,
\frac{\partial}{\partial z_{\mu}} \, \psi_{-\rho}^+ (z) \, Q_{\mu}
, \quad
\dirdi^+ \, \psi_{\rho}^+ (z) \, = \, 0 \, = \,
\frac{\partial}{\partial z_{\mu}} \, \psi_{-\rho} (z) \, Q_{\mu}^+
, \quad \rho > 0 ,
\end{equation}
and the conjugation law
\begin{equation}
\label{eq2.19new}
(\psi_{-\rho}^+ (\overline{z}))^* \, = \,
\psi_{\rho} (z)
\, .
\end{equation}
The vacuum is annihilated by $\psi_{\rho}^{(+)}$ for positive $\rho$ and the modes
$\phi_{\{ n , m \}} (z)$ of (\ref{eq1.2}) are related to $\psi_{\rho}$ by:
\begin{equation}
\label{eq2.20new}
\psi_{-\ell - \frac{3}{2}} (z) \, = \, \phi_{\{ 0 , \ell \}} (z)
, \quad
\dirz^+ \, \psi_{\ell + \frac{3}{2}} (z) \, = \, \phi_{\{-\ell - 2 , \ell + 1\}} (z)
\, .
\end{equation}
We can write
\begin{equation}
\label{eq2.21new}
\psi^{e} (z) = Y \left(\raisebox{10pt}{\hspace{-2pt}}\right.
\VL e ; \frac{3}{2} ; \frac{1}{2} , 0 \RA , z
\left.\raisebox{10pt}{\hspace{-2pt}}\right)
\ \, \text{where} \ \,
\VL e ; \frac{3}{2} ; \frac{1}{2} , 0 \RA =
\psi_{-\frac{3}{2}}^e \rvac
, \ \,
e = \pm \, (\psi^- \equiv \psi) ,
\end{equation}
the minimal energy state vectors $\VL \pm ; \frac{3}{2} ; \frac{1}{2} , 0 \RA$
being conjugate 2-component spinors.

\msection{Temperature equilibrium states and mean va\-lues}{sec3}

\msubsection{Elliptic Gibbs correlation functions}{ssec3.1}

{\it Finite temperature equilibrium (Gibbs) state correlation functions}
of 2-dimensio\-nal chiral vertex operators are (doubly periodic) elliptic functions
in the conformal time variables $\zeta$, while the characters of the corresponding
vertex algebra representations exhibit modular invariance properties \cite{Zh}.
We summarize in what follows the results of \cite{NT03} where similar properties
are established for higher even $D$.

Energy mean values in some higher dimensional models exhibit modular properties
(similar to those in a rational 2D CFT) for a ``renormalized'' (shifted) Hamiltonian
\beq\label{CH}
\CH \, := \, H \, + \, E_0
\eeq
with appropriate nonzero vacuum energy $E_0$.
Noting that the conformal Hamiltonian $\CH$ (as well as $H$)
has a bounded below discrete spectrum
of finite degeneracy, it is natural to assume the existence of the
{\it partition function}
\begin{equation}
\label{eq3.1new}
Z (\tau) \, = \, \TRACE_\VA \, q^{\CH}
\, , \quad
q \, = \, e^{2\pi i \tau}
\, , \quad
{\rm Im} \, \tau \, > \, 0 \ (\vert q \vert < 1)
\, ,
\end{equation}
as well as of all traces of the type $\TRACE_\VA (Aq^{\CH})$ where $A$ is
any polynomial in the local GCI fields. This assumption is satisfied in the theory
of free massless fields as we shall see shortly.

In order to display the periodicity properties of correlation functions
it is convenient to use the {\it compact picture} fields $\phi (\zeta , u)$
(of dimension $d_{\phi}$) related to the above $z$-picture fields by
\begin{equation}
\label{eq3.2new}
\phi (\zeta , u) \, = \, e^{2\pi i d_{\phi} \zeta} \, \phi (e^{2\pi i \zeta} \, u)
\quad \Rightarrow \quad
q^{\CH} \, \phi (\zeta , u) \rvac \, = \, q^{E_0} \, \phi (\zeta + \tau , u) \rvac
\, ; \
\end{equation}
in particular,
\beq\label{1t}
\phi (\zeta+1 , u) \, = \, \left( -1 \right)^{2d_{\phi}} \,
\phi (\zeta+1 , u)
\, . \
\eeq
We shall rewrite the expansion (\ref{eq1.2}) in the compact picture as:
\beq\label{cp_exp}
\phi (\zeta,u) \, = \, \mathop{\sum}\limits_{n \, \in \, \Z} \,
\phi_{-nm} (u) \, e^{2\pi i n\zeta}
\, , \quad
\left[ \hspace{1pt} H , \phi_{nm}(u) \hspace{1pt} \right] \, = \,
-n \, \phi_{nm} (u)
\eeq
$\phi_{nm}(u)$ being homogeneous harmonic polynomial of degree $m$.

We shall sketch the main properties of
{\it Gibbs (temperature) correlation functions} (studied in \cite{NT03})
using the simplest non-trivial example of the 2-point function
\begin{equation}
\label{eq3.3new}
w_q (\zeta_{12} ; u_1 , u_2) =
\La \phi (\zeta_1 , u_1) \, \phi^* (\zeta_2 , u_2) \Ra_q :=
\frac{\TRACE_\VA
(\phi (\zeta_1 , u_1) \, \phi^* (\zeta_2 , u_2) \, q^{\CH})}{Z(\tau)}
\end{equation}
of the compact picture local field $\phi \left( \zeta, u \right)$,
where, due to $\CH$-invariance of the right hand side, $w_q$ only depends on the
difference $\zeta_{12} = \zeta_1 - \zeta_2$ of the conformal time variables.
Our terminology (``temperature means'') is justified by the fact that the
(positive!) imaginary part of $\tau$ is identified with the inverse absolute
temperature. More precisely, as we have defined (according to (\ref{eq2.1new}) and
(\ref{equ2.8})) the conformal Hamiltonian $\CH$ to be dimensionless we should set
\begin{equation}
\label{eq3.4new}
2\pi \, {\rm Im} \, \tau = \frac{h\nu}{kT} \, ,
\end{equation}
$h\nu$ being the Planck energy quantum, $k$, the Boltzmann constant.

The cyclic property of the trace and the second equation (\ref{eq3.2new}) imply the
{\it Kubo-Martin-Schwinger (KMS) boundary condition}:
\begin{equation}
\label{eq3.5new}
\La \phi (\zeta_1 , u_1) \, \phi^* (\zeta_2 + \tau , u_2) \Ra_q \, = \,
\La \phi^* (\zeta_2 , u_2) \, \phi (\zeta_1 , u_1) \Ra_q \, .
\end{equation}
If the 2-point function is meromorphic and $\Z_2$ symmetric under
permutation of the factors (according to (\ref{1t}) and (\ref{loc}))
then $w_q$ (\ref{eq3.3new}) is elliptic in $\zeta_{12}$ satisfying
\begin{equation}
\label{eq3.6new}
w_q (\zeta_{12} \hspace{-1pt} + \hspace{-1pt} 1 ; u_1 , u_2) \, = \,
(-1)^{2 d_{\phi}} \, w_q (\zeta_{12} ; u_1 , u_2) \, = \,
\left( -1 \right)^{2 d_{\phi}} \,
w_q (\zeta_{12} \hspace{-1pt} + \hspace{-1pt} \tau ; u_1 , u_2)
,
\end{equation}
as a consequence of the KMS condition (\ref{eq3.5new}).
If $\phi$ is a generalized free field then the KMS condition
on the modes gives
\beqa\label{eqn3.9n}
&
\La
\phi_{n_1m_1} \left( u_1 \right)
\sphi_{n_2m_2} \left( u_2 \right)
\Ra_{q}
\, = & \nn & = \,
\frac{\raisebox{3pt}{\(
q^{n_2}
\)}}{\raisebox{-5pt}{\(
1-\left( -1 \right)^{2d_{\phi}} q^{n_2}
\)}}
\lvac \left[\raisebox{9pt}{\hspace{-2pt}}\right.
\phi_{n_1m_1} \left( u_1 \right),\,
\sphi_{n_2m_2} \left( u_2 \right) \,
\left.\raisebox{9pt}{\hspace{-2pt}}\right]_{\mp} \rvac
\, &
\eeqa
($\sphi_{nm} \left( u \right)$ stand for the modes of the conjugated
field $\phi^* \left( \zeta,u \right)$)
so that if we expand \(\txfrac{q^{n_2}}{
1-\left( -1 \right)^{2d_{\phi}} q^{n_2}}\) in a progression
for \(|q| < 1\) and then take the corresponding sums~(\ref{n1.3})
over the modes we will obtain (\cite{NT03} Theorem~4.1):
\beq\label{eqn3.10n}
w_q (\zeta_{12} ; u_1 , u_2) \, = \,
\mathop{\sum}\limits_{k \, = \, -\infty}^{\infty}
\left( -1 \right)^{2 \hspace{1pt} k \hspace{1pt} d_{\phi}} \,
w_0 \left( \zeta_{12} \hspace{-1pt} + \hspace{-1pt} k\,\tau \hspace{1pt} ; \hspace{1pt}
u_1 , u_2 \right)
\, ,
\eeq
where \(w_0 (\zeta_{12} ; u_1 , u_2)\)
is the vacuum (\(q=0\)) 2-point function.
Indeed, the progression term \(\left(
\left( -1 \right)^{2d_{\phi}} \hspace{-1.5pt} q^{|n_2|}
\right)^{\pm k}\)\gvspc{15pt}
multiplying the vacuum expectations
will produce the term
\(\left( -1 \right)^{2 \hspace{1pt} k \hspace{1pt} d_{\phi}} \hspace{-1.5pt}
w_0 \left( \zeta_{12} \hspace{-1pt} + \hspace{-1pt} k\,\tau \hspace{1pt} ; \hspace{1pt}
u_1 , u_2 \right)\) in the sum~(\ref{eqn3.10n})
(see Remark~\ref{rm3.1n} below).

In particular, for a 4-dimensional massless scalar field $\varphi (\zeta , u)$
with 2-point function
\beq
\label{eq3.7new}
\lvac \varphi (\zeta_1 , u_1) \, \varphi (\zeta_2 , u_2) \rvac
\, = \, \frac{e^{2\pi i (\zeta_1 + \zeta_2)}}{
(e^{2\pi i \zeta_1} \, u_1 - e^{2\pi i \zeta_2} \, u_2)^2}
\, = \, \frac{-1}{4\,\sin \pi \zeta_+ \sin \pi \zeta_-} \, ,
\eeq
where
\beq
\label{eq3.8new}
\cos 2 \pi \alpha \, := \, u_1 \spr u_2
\, , \quad
\zeta_{\pm} = \zeta_{12} \pm \alpha \, ,
\eeq
the Gibbs correlation function is given by:
\begin{equation}
\label{eq3.9new}
w_q (\zeta_{12} , \alpha) =
\La \varphi (\zeta_1 , u_1) \, \varphi (\zeta_2 , u_2) \Ra_q
= \frac{1}{4\pi \sin 2\pi \alpha} \, (p_1 (\zeta_+ , \tau) -
p_1 (\zeta_- , \tau)) \, ,
\end{equation}
\beq\label{eq3.10new}
p_1 (\zeta , \tau)
\, = \, \lim_{M \to \infty} \lim_{N \to \infty} \sum_{m=-M}^M \sum_{n=-N}^N
\frac{1}{\zeta + m\tau + n}
\, . \
\eeq
(Note that $p_1 (\zeta , \tau)$ is not elliptic but the difference (\ref{eq3.9new})
is.) In order to derive (\ref{eq3.9new}) from (\ref{eqn3.10n})
one should use the identity
$$
\frac{-1}{\sin \pi \zeta_+ \sin \pi \zeta_-} \, = \,
\frac{1}{\sin 2 \pi \alpha} \, (\cot \pi \zeta_+ - \cot \pi \zeta_-) \, ,
$$
and the Euler formula
\begin{equation}
\label{eq3.14new}
\pi \cot \pi \zeta \, = \, \lim_{N \to \infty} \sum_{n=-N}^N \frac{1}{\zeta + n}
\, .
\end{equation}

\begin{mremark}\label{rm3.1n}
The mode expansion (\ref{eq2.11new}) of the compact picture field
$\varphi(\zeta,u)$ reads:
\beq\label{e3.17n}
\varphi (\zeta , u) \, = \,
\sum_{n \in \Z} \varphi_n (u) \, e^{-2\pi i n \zeta}
\quad
(\hspace{1pt} \varphi_n (u) \rvac = 0 \quad \text{for} \quad n \geqslant 0 \hspace{1pt})
\, .
\eeq
(comparing with (\ref{cp_exp}) we have
\(\phi_{nm}(u) \equiv \delta_{m+1,|n|} \, \varphi_n(u)\));
then Eq.~(\ref{eqn3.9n}) takes the form
\beq\label{e3.18n}
\La \varphi_{-m} (u_1) \, \varphi_n (u_2) \Ra_q \
\, = \, \delta_{mn} \, \frac{q^n}{1-q^n} \,
\frac{\sin 2\pi n \alpha}{\sin 2\pi\alpha}
\, . \
\eeq
We thus obtain the following $q$--expansion of $w_q(\zeta,\alpha)$:
\beq\label{e3.19n}
w_q (\zeta,\alpha) \, = \,
w_0 (\zeta,\alpha) \, + \,
2 \mathop{\sum}\limits_{n \, = \, 1}^{\infty} \,
\frac{q^n}{1-q^n} \, \frac{\sin 2\pi n \alpha}{
\sin 2\pi\alpha} \hspace{1pt} \cos 2\pi n \zeta
\eeq
which can be derived also from Eq.~(\ref{eq3.9new}) and the $q$--expansion
\beq\label{e3.20n}
p_1 (\zeta,\tau)
\, = \, \pi \cot \pi \zeta + 4\pi \sum_{n=1}^{\infty}
\frac{q^n}{1-q^n} \, \sin 2 \pi n \zeta \, .
\eeq
\end{mremark}

In order to compute the Gibbs 2-point function of a Weyl field we first write its
vacuum Wightman function in the form
\begin{eqnarray}
\label{eq3.15new}
&& \lvac \psi (\zeta_1 , u_1) \, \psi^+ (\zeta_2 , u_2) \rvac \, = \,
\frac{i}{8 \sin \pi \zeta_- \sin \pi \zeta_+}
\left( \frac{{\dirv}^+}{\sin \pi \zeta_-} +
\frac{\bar{\dirv}^+}{\sin \pi \zeta_+} \right) \nonumber \\
&& = \, \frac{i}{8 \sin 2 \pi \alpha} \biggl\{ {\dirv}^+
\left( \frac{\cos \pi \zeta_-}{\sin^2 \pi \zeta_-} -
\frac{\cot 2 \pi \alpha}{\sin \pi \zeta_-} +
\frac{1}{\sin 2 \pi \alpha \sin \pi \zeta_+} \right) \nonumber \\
&& \hspace{12pt} \,
- \, \bar{\dirv}^+ \left( \frac{\cos \pi \zeta_+}{\sin^2 \pi \zeta_+}
+ \frac{\cot 2 \pi \alpha}{\sin \pi \zeta_+}
- \frac{1}{\sin 2 \pi \alpha \sin \pi \zeta_-} \right) \biggl\}
\end{eqnarray}
where $v$ and $\overline{v}$ are conjugate complex isotropic 4-vectors determined
for non-collinear $u_1 , u_2$ from the equations
\begin{equation}
\label{eq3.16new}
u_1 = e^{i\pi\alpha} \, v + e^{-i\pi\alpha} \, \overline{v}
, \quad
u_2 = e^{-i\pi \alpha} \, v + e^{i\pi\alpha} \, \overline{v}
\quad
(u_1^2 = u_2^2 = 1 \ \, \Rightarrow \ \, 2v \spr \overline{v} = 1).
\end{equation}
Then Eq.~(\ref{eqn3.10n}) gives
\begin{eqnarray}
\label{eq3.17new}
&&
\hspace{-30pt}
\La \psi (\zeta_1 , u_1) \, \psi^+ (\zeta_2 , u_2) \Ra_q \, = \,
\frac{i}{8\pi \sin 2\pi \alpha} \biggl\{ {\dirv}^+
\Bigl( p_2^{11} (\zeta_- , \tau) -
\cot 2 \pi \alpha \, p_1^{11} (\zeta_- , \tau)
\nonumber \\
&&
\hspace{-30pt}
\hspace{12pt} +
\frac{p_1^{11} (\zeta_+ , \tau)}{\sin 2 \pi \alpha} \Bigl)
- \,
\bar{\dirv}^+ \left( p_2^{11} (\zeta_+ , \tau) +
\cot 2 \pi \alpha \, p_1^{11} (\zeta_+ , \tau) -
\frac{p_1^{11} (\zeta_- , \tau)}{\sin 2 \pi \alpha} \right) \biggl\} .
\quad
\end{eqnarray}
Here we are using the functions $p_k^{\kappa , \lambda} (\zeta , \tau)$
given by the Eisenstein series
\begin{equation}
\label{eq3.18new}
p_k^{\kappa , \lambda} (\zeta , \tau) \, = \,
\sum_{m,n \in \Z} \frac{(-1)^{\kappa m +
\lambda n}}{(\zeta + m\tau + n)^k}
\, , \quad
\kappa , \lambda = 0,1
\end{equation}
which are not absolutely convergent for $k \leqslant 2$ and hence require specifying
order of limits; in particular,
\begin{equation}
\label{eq3.19new}
p_1^{11} (\zeta , \tau) \, = \,
\pi \sum_{n \in \Z} \frac{(-1)^n}{\sin \pi (\zeta + n \tau)}
\, , \quad
p_2^{11} (\zeta , \tau) \, = \,
- \frac{\partial}{\partial \zeta} \, p_1^{11} (\zeta , \tau) \, .
\end{equation}
The functions $p_k^{\kappa , \lambda}$ are defined (in Appendix A of \cite{NT03})
to satisfy
$$
p_k^{\kappa , \lambda} (\zeta + 1 , \tau) =
(-1)^{\lambda} \, p_k^{\kappa , \lambda} (\zeta , \tau)
\, , \quad p_k^{\kappa , \lambda} (\zeta + \tau , \tau) =
(-1)^{\kappa} \, p_k^{\kappa , \lambda} (\zeta , \tau)
$$
for \(k + \kappa + \lambda > 1\).

\msubsection{Energy mean values as (combinations of) modular forms}{ssec3.2}

Knowing the temperature 2-point function one can compute by
differentiating in $\zeta$ and equating the two points (after subtracting the
singularity for coinciding arguments) the energy mean value in the corresponding
equilibrium state.
This procedure is based on the assumption that the energy momentum tensor
occurs in the OPE of the product of two basic fields.
We shall follow a more specific procedure ({\it cf.} \cite{DK}) applicable
whenever there is a Fock space realization of the state space
$\VA$ ({\it e.g.}, in the bosonic case, whenever $\VA$ can be
viewed as a ``symmetric exponent'' of a 1-particle space). To this end one only
needs the degeneracy of eigenvalues of the conformal Hamiltonian $\CH$ in the
1-particle subspace. Rather than describing the general method we shall illustrate
it in two typical examples: a free scalar field in dimension $D = 2d_0 + 2$
({\it cf.} (\ref{eq2.10new})) and the free Weyl field in four dimension.

The 1-particle state space in the theory of a free hermitean scalar field $\varphi$
in dimension $D$ is the direct sum of spaces of homogeneous harmonic polynomials
$\{ \varphi_{-n} (z) \rvac \}$ of degree $n-d_0$ ($n= d_0 , d_0 + 1 , \ldots$).
The dimension of each such eigenspace of $\CH$ is $d (n-d_0 , D)$, where
\begin{eqnarray}
\label{eq1.8new}
&&d(m,D) \, = \,
\begin{pmatrix} m+D-1 \\ D-1 \end{pmatrix} - \begin{pmatrix} m+D-3 \\ D-1
\end{pmatrix} , \nonumber \\
&&\text{\it i.e.} \quad
d(n-d_0 , D) \, = \, \frac{2}{(2d_0)!} \prod_{i=0}^{d_0 - 1} (n^2 - i^2)
\end{eqnarray}
is the dimension of the space of homogeneous harmonic polynomials of degree $m$
in $\C^D$ which carries an irreducible representation of
$\Spin (D)$;
hence,
\begin{equation}
\label{eq3.20new}
\La \CH \Ra_q^{(d_0)} \, = \,
E(d_0) + \sum_{n=d_0}^{\infty} \frac{2}{(2 d_0)!} \, n^2
\ldots
[n^2 - (d_0 - 1)^2] \, \frac{nq^n}{1-q^n} \, .
\end{equation}
For $D=4$ ($d_0 = 1$) the energy mean value is a modular form (of weight 4),
\begin{equation}
\label{eq3.21new}
\La \CH \Ra_q^{(1)} \, = \, G_4 (\tau)
\quad \text{for} \quad
E(1) \, = \, \lvac \CH \rvac \, = \, - \frac{B_4}{8} \, = \, \frac{1}{240}
\end{equation}
where we are using the notation and conventions of \cite{Za}:
\begin{eqnarray}
\label{eq3.22new}
&& G_{2k} (\tau) \, := \,
- \frac{B_{2k}}{4k} + \sum_{n \, = \, 1}^{\infty} \,
\frac{n^{2k-1}}{1-q^n} \ q^n
\nonumber \\
&& \left(\raisebox{12pt}{\hspace{-2pt}}\right.
= \frac{(2k-1)!}{(2\pi i)^{2k}}
\left\{\raisebox{14pt}{\hspace{-2pt}}\right.
\sum_{n=1}^{\infty} \frac{1}{n^{2k}} + \sum_{m \, = \, 1}^{\infty} \,
\sum_{n \in \Z} (m\tau + n)^{-2k}
\left.\raisebox{14pt}{\hspace{-2pt}}\right\}
\quad \text{for} \quad k \geqslant 2
\left.\raisebox{12pt}{\hspace{-2pt}}\right)
\qquad
\end{eqnarray}
(the Eisenstein series in the braces being only absolutely convergent
for $k \geqslant 2$).

\medskip

\begin{mremark}\label{rm3.1}
We note that for pure imaginary $\tau$ related to the absolute temperature $T$ by
(\ref{eq3.4new}), multiplying $\CH$ in (\ref{eq3.21new}) by $h\nu$ in order to
express it in units of energy we find for the $n$-th term in the expansion of
$G_4$ (\ref{eq3.21new}) the {\it Planck distribution law}
$$
\frac{n^3 \, h \, \nu}{e^{\frac{hn\nu}{kT}} - 1}
$$
(up to an overall factor, $\txfrac{8\pi \nu^2}{c^3}$,
independent of $n$ and $T$).
\end{mremark}

For $D=6$ we find from (\ref{eq3.20new}) a linear combination of modular forms of
weight 6 and 4:
\begin{equation}
\label{eq3.24new}
\La \CH \Ra_q^{(2)} \, = \,
\frac{1}{12} \, [G_6 (\tau) - G_4 (\tau)] \, = \,
\frac{-31}{12 \times 7!} + \frac{18 \, q^3}{1-q^3}
+ \frac{80 \, q^4}{1-q^4} + \cdots
\end{equation}
(the coefficients to $q^n$ for \(n > 0\) being thus positive integers).
The functions $G_{2k} (\tau)$ for \(k > 1\) are {\it modular forms of weight}
$2k$ (and level 1):
\begin{eqnarray}
\label{eq3.25new}
&& (c\tau + d)^{-2k} \, G_{2k} \left( \frac{a\tau + b}{c\tau + d} \right) \, = \,
G_{2k} (\tau)
\nonumber \\
&& \text{for} \quad \gamma \, := \,
\begin{pmatrix} a &b \\ c & d \end{pmatrix} \in \Gamma (1) :=
\mathit{SL} (2,\Z) \quad (k = 2,3,\ldots) \, .
\end{eqnarray}
There is no level 1 modular form of weight 2; the function $G_2$ (given by the
first equation (\ref{eq3.22new})) transforms inhomogeneously under $\Gamma (1)$:
\begin{equation}
\label{eq3.26new}
\frac{1}{(c\tau + d)^2} \, G_2 \left( \frac{a\tau + b}{c\tau + d} \right) \, = \,
G_2 (\tau) + \frac{ic}{4\pi (c\tau + d)} \, .
\end{equation}

The 1-particle state space for a complex Weyl field is the direct sum of
positive and negative charge states of energy
\(\left( n + \txfrac{1}{2} \right)\) (\(n = 1,2,\ldots\))
of equal multipli\-city.
The Weyl equation (\ref{eq2.18new}) implies that the positive charge energy
eigenspace spanned by
\(\left\{\raisebox{10pt}{\hspace{-2pt}}\right.
\psi_{-n - \frac{1}{2}}^+ (z) \rvac
\left.\raisebox{10pt}{\hspace{-2pt}}\right\}\) carries the irreducible
\(\Spin (4) = SU(2) \times SU(2)\) representation
\(\left(\raisebox{10pt}{\hspace{-2pt}}\right.
\txfrac{n}{2} , \txfrac{n-1}{2}
\left.\raisebox{10pt}{\hspace{-2pt}}\right)\) of dimension $n(n+1)$.
The negative charge energy eigenspace spanned by
\(\left\{\raisebox{10pt}{\hspace{-2pt}}\right.
\psi_{-n - \frac{1}{2}} (z) \rvac \left.\raisebox{10pt}{\hspace{-2pt}}\right\}\)
has the same dimension.
As a result, the mean energy in an equilibrium temperature state is given by
\begin{eqnarray}
\label{eq3.27new}
&& \La \CH \Ra_q \, = \,
E_0 + \sum_{n \, = \, 1}^{\infty} \, (2n+1) \, \frac{n(n+1) \,
q^{n + \frac{1}{2}}}{1 + q^{n + \frac{1}{2}}}
\nonumber \\
&& = \, \frac{1}{4}
\left\{\raisebox{10pt}{\hspace{-2pt}}\right.
G_4 \left(\raisebox{10pt}{\hspace{-2pt}}\right.
\frac{\tau + 1}{2} \left.\raisebox{10pt}{\hspace{-2pt}}\right)
- 8 \, G_4 (\tau) - G_2 \left(\raisebox{10pt}{\hspace{-2pt}}\right.
\frac{\tau + 1}{2} \left.\raisebox{10pt}{\hspace{-2pt}}\right)
+ 2 \, G_2 (\tau)
\left.\raisebox{10pt}{\hspace{-2pt}}\right\} \, .
\end{eqnarray}
To prove the first equation one again uses the KMS condition combined with the
canonical {\it anticommutation relations} for the $\psi$ modes.
The second is derived from the identities
\begin{equation}
\label{eq3.28new}
n(n+1) \, = \,
\frac{1}{4} \, \left[\raisebox{9pt}{\hspace{-2pt}}\right.
(2n+1)^2 - 1 \left.\raisebox{9pt}{\hspace{-2pt}}\right]
\, , \quad
q \left(\raisebox{10pt}{\hspace{-2pt}}\right. \frac{\tau + 1}{2}
\left.\raisebox{10pt}{\hspace{-2pt}}\right) \, = \,
-q^{1/2} \quad (\text{for} \ \ q \, \equiv \, q (\tau)) \, .
\end{equation}
We note that although $G_2 (\tau)$ is not a modular form the difference
\begin{equation}
\label{eq3.29new}
F(\tau) \, := \, 2 \, G_2 (\tau) - G_2 \left(\raisebox{10pt}{\hspace{-2pt}}\right.
\frac{\tau + 1}{2} \left.\raisebox{10pt}{\hspace{-2pt}}\right) \, ,
\end{equation}
equal to the energy distribution of a chiral Weyl field (in 2-dimensional conformal
field theory), is a weight 2 form with respect to the index 2 subgroup
\(\Gamma_{\theta} \subset \Gamma (1)\) generated by
\(S = \begin{pmatrix} 0 &-1 \\ 1 &0 \end{pmatrix}\) and
\(T^2 = \begin{pmatrix} 1 &2 \\ 0 &1 \end{pmatrix}\).

\msection{General 4-point functions}{sec4}

\msubsection{Strong locality and energy positivity imply rationality}{ssec4.1}

As already stated in Sect.~\ref{ssec2.1} GCI and Wightman axioms imply strong
locality (\ref{loc}) for vertex operators.
In other words, for any pair of conjugate (Bose or Fermi) fields $\psi (z_1)$ and
$\psi^* (z_2)$ there is a positive integer $N_{\psi}$ such that
\begin{equation}
\label{eq2.1}
\rho_{12}^N \left\{\raisebox{10pt}{\hspace{-2pt}}\right.
\psi (z_1) \, \psi^* (z_2) - \varepsilon_{\psi} \, \psi^* (z_2) \,
\psi (z_1) \left.\raisebox{10pt}{\hspace{-2pt}}\right\} \, = \, 0
\quad \text{for} \quad N \geqslant N_{\psi}
\quad
(\rho_{12} \, := \, z_{12}^{\, 2})
\end{equation}
where $\varepsilon_{\psi} (= \pm 1)$ is the fermion parity of $\psi$. If $\psi$
transforms under an {\it elementary local field representation} of the spinor
conformal group $SU (2,2)$ (see \cite{M77,T85}) -- {\it i.e.} one induced by a
$(2j_1 + 1)(2j_2 + 1)$ dimensional representation $(d;j_1 , j_2)$ of the maximal
compact subgroup $S(U(2) \times U(2))$ of $SU (2,2)$ ($d$ being the $U(1)$
character coinciding with the scale dimension), then
\begin{equation}
\label{eq2.2}
N_{\psi} \, = \, d + j_1 + j_2
\, , \quad
\varepsilon_{\psi} \, = \, (-1)^{2j_1 + 2j_2} \, = \, (-1)^{2d} \, .
\end{equation}
It follows that for any $n$-point function of GCI local fields and for large
enough $N \in \N$ the product
\begin{equation}
\label{eq2.3}
F_{1 \ldots n} (z_1 , \ldots , z_n) \, := \,
\left(\raisebox{14pt}{\hspace{-2pt}}\right.
\prod_{1 \, \leqslant \, i \, < \, j \, \leqslant \, n} \,
\rho_{ij} \left.\raisebox{14pt}{\hspace{-2pt}}\right)^{\hspace{-2pt} N}
\lvac \phi_1 (z_1) \ldots \phi_n (z_n)
\rvac
\end{equation}
$(\rho_{ij} = z_{ij}^2 \equiv (z_i - z_j)^2)$, is $\Z_2$ symmetric
under any permutation of the factors within the vacuum expectation value. Energy
positivity, on the other hand, implies that $\lvac \phi_1 (z_1) \ldots
\phi_n (z_n) \rvac$, and hence $F_{1 \ldots n} (z_1 , \ldots , z_n)$ do not
contain negative powers of $z_n^2$. It then follows from the symmetry and the
homogeneity of $F_{1 \ldots n}$ that it is
a polynomial in all $z_i^{\mu}$. Thus the (Wightman) correlation functions are
rational functions of the coordinate differences. (See for more detail \cite{NT03};
an equivalent Minkowski space argument based
on the support properties of the Fourier transform of (the $x$-space counterpart
of) (\ref{eq2.3}) is given in \cite{NT01}.)

Rationality of correlation functions implies that all dimensions of GCI fields
should be integer or half integer depending on their spin, more precisely, that
sums like $N_{\psi}$ (\ref{eq2.2}) should be integer. This condition is,
however, not sufficient for rationality even of 3-point functions.

\medskip

\begin{mobservation} \label{ob4.1}
The necessary and sufficient condition for
the existence of a GCI $3$-point function \(\lvac \phi_1 (z_1) \,
\phi_2 (z_2) \, \phi_3 (z_3) \rvac\) of elementary conformal fields
$\phi_i (z)$ of $S(U(2) \times U(2))$ weight $(d_i ; j_{i1} , j_{i2})$ is:
\begin{equation}
\label{eq2.5}
N_i \, := \, d_i + j_{i1} + j_{i2} \in \N
\, , \quad
\frac{1}{2} \sum_{i=1}^{3} \,
N_i \, \in \, \N
\, , \quad
\sum_{i=1}^{3} d_i \, \in \, \N \, .
\end{equation}
\end{mobservation}

The statement follows from the explicit knowledge of 3-point functions (see for
reviews \cite{TMP} \cite{T85} \cite{FP}).

In particular, there is no Yukawa type rational conformal 3-point function of a
pair of conjugate canonical (\(\, d = \txfrac{3}{2}\,\)) spinor fields and a
canonical (\(d=1\)) scalar field. Similarly, one observes that 2-point functions
of free massless fields in odd space-time dimensions are not rational and hence
cannot be GCI.

It is important for the feasibility of constructing a GCI model that the
singularities of $n$-point functions (the integer $N$ in (\ref{eq2.3})) is
majorized by that of the 2-point function ($N_{\psi}$ in (\ref{eq2.2})) whenever
Wightman positivity is satisfied.

\msubsection{General truncated 4-point function of a GCI scalar
field}{ssec4.2}

Infinitesimal (or Euclidean) conformal invariance is sufficient to determine 2-
and 3-point functions (see, {\it e.g.} \cite{TMP}). One can construct, however,
two independent conformally invariant cross-ratios out of four points,
\begin{equation}
\label{eq2.6}
s \, = \, \frac{\rho_{12} \rho_{34}}{\rho_{13} \rho_{24}}
\, , \quad
t \, = \, \frac{\rho_{14} \rho_{23}}{\rho_{13} \rho_{24}} \, ,
\end{equation}
so that a simple minded symmetry argument does not determine the 4-point
functions. GCI, on the other hand, combined with Wightman axioms, yields
rationality and thus allows to construct higher point correlation functions
involving just a finite number of free parameters. In particular, the truncated
4-point function of a hermitean scalar field $\phi$ of (integer) dimension $d$
can be written in the form (\cite{NST02} Sect.~1):
\begin{eqnarray}
\label{eq2.7}
w_4^t \, \equiv \podr w^t (z_1 , z_2 , z_3 , z_4) \, := \,
\langle 1234 \rangle
- \langle 12 \rangle \langle 34 \rangle
- \langle 13 \rangle \langle 24 \rangle
- \langle 14 \rangle \langle 23 \rangle
\nonumber \\
= \podr
\frac{(\rho_{13} \rho_{24})^{d-2}}{(\rho_{12} \rho_{23} \rho_{34}
\rho_{14})^{d-1}} \, P_d (s,t)
\, , \quad
P_d (s,t) \, = \, \sum_{i \, \geqslant \, 0 , j \, \geqslant \, 0}^{i+j \, \leqslant \, 2d-3}
c_{ij} s^i t^j
\, ,
\end{eqnarray}
where $\langle 1 \ldots n \rangle$ is a short-hand for the $n$-point function of
$\phi$:
\begin{equation}
\label{eq2.8}
\langle 1 \ldots n \rangle \, = \, \lvac \phi (z_1) \ldots \phi (z_n) \rvac
\quad
( \hspace{1pt}
\langle 12 \rangle \, = \, B_{\phi} \, \rho_{12}^{-d}
\hspace{1pt} )
\, .
\end{equation}
In writing down (\ref{eq2.7}) we have used the fact that for space-time
dimensions \(D > 2\) Hilbert space positivity implies that the degree of
singularities of the truncated $n$-point function \((n \geqslant 4)\) is strictly
smaller than the degree of the 2-point function.

Furthermore, crossing symmetry (which is a manifestation of local commutativity)
implies an $\Ss_3$ symmetry of $P_d$:
\begin{eqnarray}
\label{eq2.9}
&& s_{i \hspace{1pt} i+1} P_d (s,t) \, = \, P_d (s,t) , \, i = 1,2
\, , \quad \nonumber \\
&& s_{12} \, P_d (s,t) \, := \, t^{2d-3}
P_d \left( \frac{s}{t} , \frac{1}{t} \right)
\, , \quad
s_{23} \, P_d (s,t) \, := \, s^{2d-3}
P_d \left( \frac{1}{s} , \frac{t}{s} \right) \, , \qquad
\end{eqnarray}
$s_{ij}$ being the substitution exchanging the arguments $z_i$ and $z_j$. The
number of independent crossing symmetric polynomials $P_d$ is
$\dbrackets{\txfrac{d^2}{3}}$
(the integer part of $d^2 / 3$:
\(\dbrackets{\txfrac{d^2}{3}} = n (2d - 3n)\) for
\(3n-1 \leqslant d \leqslant 3n+1$, $n = 0,1,2,\ldots\)).

The 1-parameter family of crossing symmetric polynomials for $d=2$ is
\(P_2 (s,t) = c \, (1+s+t)\), {\it i.e.}
\begin{equation}
\label{eq2.10}
w_4^t \, = \, c \,
\left\{\raisebox{10pt}{\hspace{-2pt}}\right.
(\rho_{12} \rho_{23} \rho_{34} \rho_{14})^{-1} + (\rho_{13} \rho_{23}
\rho_{24} \rho_{14})^{-1} + (\rho_{12} \rho_{13} \rho_{24} \rho_{34})^{-1}
\left.\raisebox{10pt}{\hspace{-2pt}}\right\} \, ,
\end{equation}
thus corresponding to the sum of three 1-loop diagrams for a sum of normal
products of free massless fields:
\begin{equation}
\label{eq2.11}
\phi (z) \, = \,
\frac{1}{2} \sum_{i=1}^{N} : \varphi_i^2 (z) :
\quad
([\varphi_i (z_1) , \varphi_j (z_2)] \, = \, 0 \quad \text{for} \quad
i \neq j \, , \quad \Delta \, \varphi_i (z) \, = \, 0) \, .
\end{equation}
Indeed, it was proven in \cite{NST02} that $\phi (z)$ generates under
commutations a central extension of the infinite symplectic algebra
\(\mathit{sp} (\infty , \R)\) for \(d=2\) and that the unitary vacuum
representations of this algebra correspond to integer central charge
\(c = N \, (\in \N \,)\).
Thus, (\ref{eq2.11}) is the general form of a $d=2$ GCI field satisfying Wightman
axioms (including Hilbert space positivity) and involving a unique rank 2 symmetric
traceless tensor of dimension 4 in its OPE algebra.

The physically most attractive example, corresponding to a $d=4$ scalar field
$\Lagr (z)$ that can be interpreted as a QFT Lagrangian density, gives
rise to a 5-parameter truncated 4-point function \cite{NST03} of type
(\ref{eq2.7}) with
\def \shfta {\hspace{-26pt}}
\beqa
\shfta &
\label{eq2.12}
P_4 (s,t) \hspace{1pt} = \hspace{1pt}
\sum_{\nu \, = \, 0}^2 \, a_{\nu} J_{\nu} (s,t) +
st \hspace{1pt} \left[\raisebox{10pt}{\hspace{-2pt}}\right.
b (Q_1 (s,t) - 2Q_2 (s,t)) + c \, Q_2 (s,t)
\left.\raisebox{10pt}{\hspace{-2pt}}\right] \hspace{-1pt} ,
& \\ \shfta &
\label{eq2.13}
J_0 (s,t) \hspace{1pt} := \hspace{1pt} s^2 (1+s) + t^2 (1+t) + s^2 t^2 (s+t) ,
& \\ \shfta &
\label{eq2.14}
J_1 (s,t) := s(1 \hspace{-1pt} - \hspace{-1pt} s)
(1 \hspace{-1pt} - \hspace{-1pt} s^2) + t(1 \hspace{-1pt} - \hspace{1pt} t)
(1 \hspace{-1pt} - \hspace{-1pt} t^2) + st
\left[\raisebox{9pt}{\hspace{-2pt}}\right.
(s \hspace{-1pt} - \hspace{-1pt} t)(s^2 \hspace{-1pt} - \hspace{-1pt} t^2)
\hspace{-1pt} - \hspace{-1pt}
2Q_1 \left.\raisebox{9pt}{\hspace{-2pt}}\right] \hspace{-1pt} ,
& \mgvspc{10pt} \\ \shfta &
\label{eq2.15}
J_2 (s,t) := (1+t)^3 [(1+s-t)^2 - s] - 3s \, (1-t) + s^3 \,
\left[\raisebox{9pt}{\hspace{-2pt}}\right.
(1+t-s)^2 - t \left.\raisebox{9pt}{\hspace{-2pt}}\right] \hspace{-1pt} ,
& \mgvspc{10pt} \\ \shfta &
\label{eq2.16}
Q_1 (s,t) \, := \, 1 + s^2 + t^2 \, , \quad Q_2 (s,t) := s+t+st
\, , & \mgvspc{10pt} \nn \shfta &
Q_1 (s,t) - 2Q_2 (s,t) \, = \, (1-s-t)^2 - 4st \, .
&
\eeqa
As we shall see in the next section, the $J_{\nu}$ polynomials correspond
to the twist 2 fields' contribution to the OPE, symmetrized in an appropriate way,
while the terms $st \, Q_j$ correspond to twist 4 and higher contributions.

The above choice of basic $\Ss_3$ symmetric polynomials is not
accidental: it is essentially determined by its relation to the ``partial wave''
expansion of $w_4^t$ to be displayed in the next section.

\msection{OPE in terms of bilocal fields}{sec5}

\msubsection{Fixed twist fields. Conformal partial wave expansion}{ssec5.1}

From now on we shall study the GCI theory of the above \(d=4\) hermitean
scalar field $\Lagr(z)$ (in \(D=4\) space-time dimensions).

The infinite series of
local tensor fields appearing in
the OPE of $\Lagr (z_1) \, \Lagr (z_2)$ can be organized
into an infinite sum of scalar fields depending on both arguments $z_1$ and $z_2$:
\beq
\label{eq3.1}
\Lagr (z_1) \, \Lagr (z_2) \, = \,
\frac{B}{\rho_{12}^4} \, + \, \sum_{\kappa \, = \, 1}^{\infty} \,
\rho_{12}^{\kappa - 4} \, V_{\kappa} (z_1 , z_2)
\, .
\eeq
By definition, the field $V_{\kappa} \left( z_1,z_2 \right)$ is
regular for coinciding arguments (see Remark~\ref{rm5.1} below)
and its Taylor expansion in $z_{12}$ involves only
{\it twist $2\kappa$ tensor fields} and their derivatives.
More precisely, it can be written in the form
({\it cf.} \cite{FGGP} \cite{DMPPT} \cite{TMP} \cite{DO01} \cite{NST02})
\begin{equation}
\label{eq3.2}
V_{\kappa} (z_1 , z_2) \, = \,
\sum_{\ell \, = \, 0}^{\infty} \, C_{\kappa \ell} \,
K_{\kappa \ell} (z_{12} \spr \di' ,\, \rho_{12} \, \Delta') \,
O_{2\kappa, \ell} (z_2; z_{12})
\, .
\end{equation}
Here $O_{2\kappa , \ell} (z;w)$ are (contracted) symmetric traceless tensor fields,
\beq
\label{eq3.3}
O_{2\kappa , \ell} (z;w) = O_{2\kappa}^{\mu_1 \ldots \mu_{\ell}} (z) \, w_{\mu_1}
\ldots w_{\mu_{\ell}}
\quad
(\text{tracelessness} \ \, \Leftrightarrow \ \,
\Delta_w \, O_{2\kappa , \ell} (z;w) = 0)
,
\eeq
of scale dimension $d_{\kappa , \ell} = 2\kappa + \ell$ ({\it i.e.} of {\it fixed
twist} $d_{\kappa \ell} - \ell = 2\kappa$); $K_{\kappa\ell} \left( t_1,t_2 \right)$
is the Taylor series in $t_{1,2}$ of the analytic function
\beq\label{eq5.4n}
K_{\kappa,\ell} \left( t_1,t_2 \right) \, := \,
\mathop{\sum}\limits_{n \, = \, 0}^{\infty} \,
\int_0^1 \,
\frac{
\left[ \hspace{1pt} \alpha \hspace{1pt} (1-\alpha) \hspace{1pt} \right]^{
\ell \hspace{1pt} + \hspace{1pt} \kappa \hspace{1pt} + \hspace{1pt}
n \hspace{1pt} - \hspace{1pt} 1}
e^{\alpha \, t_1} \, \left( -t_2 \right)^n}{
4^n \,B(\ell + \kappa , \ell + \kappa)\, n! \, (2\ell + 2\kappa -1)_n}
\ d\alpha
\,
\eeq
(\(\left( a \right)_n :=
\txfrac{\Gamma \left( a+n \right)}{\Gamma \left( a \right)}\))
and $t_1$, $t_2$ are substituted in~(\ref{eq3.2}) by the operators
$z_{12} \spr \di'$ and $\rho_{12} \Delta'$ (resp.), where
\beq\label{eq5.5n}
\di'_{\mu} \, O_{2\kappa ,\, \ell}
(z_2 ; z_{12})
\, := \, \frac{\di O_{2\kappa , \ell} (z;w)}{\di z^{\mu}}
\hspace{1pt}\vrestr{10pt}{z = z_2,\, w = z_{12}}
\, , \quad
\Delta' \, := \, \di' \spr \di'
\, . \
\eeq
The integrodifferential operator with kernel $K_{k\ell}$ in the right
hand side of (\ref{eq3.2}) is defined to transform the 2-point function of
$O_{2\kappa , \ell}$ into the 3-point function
\beq
\label{eq3.4}
\lvac V_{\kappa} (z_1 , z_2) O_{2\kappa , \ell} (z_3 ; w)
\rvac = A_{\kappa \ell} \,
\frac{
1
}{
(\rho_{13} \rho_{23})^{\kappa}}
\left( \hspace{-2pt} \left(
\frac{z_{13}}{\rho_{13}} \hspace{-1pt} - \hspace{-1pt} \frac{z_{23}}{\rho_{23}}
\right) \hspace{-2pt} \spr w \right)^{\hspace{-1pt} \ell}
\ \, \text{for} \ \, w^2 = 0
\eeq
(see \cite{DO01} and \cite{NST03} Sect.~3, where more general OPE -- for any scale
dimension $d$ and for complex fields are considered, and the kernel
$K_{\kappa \ell}$ is written down).
For real $\Lagr$, due to the strong locality,
$V_{\kappa} \left( z_1,z_2 \right)$ should be symmetric:
\beq\label{eq3.2a}
V_{\kappa} (z_1 , z_2) \, = \, V_{\kappa} (z_2 , z_1)
\, ; \
\eeq
it then follows that only even rank tensors (even $\ell$) appear in the
expansion~(\ref{eq3.2}).
Since every field $V_{\kappa} \left( z_1,z_2 \right)$ corresponds to
a conformally invariant part of the OPE of
\(\Lagr \left( z_1 \right) \Lagr \left( z_2 \right)\)
then $V_{\kappa}$ should be a conformally invariant scalar field of dimension
$\left( \kappa,\kappa \right)$,
because of the factor \(\rho_{12}^{\kappa-4}\) in~(\ref{eq3.1}).

\begin{mremark}\label{rm5.1}
The condition of regularity of a field $V(z,w)$ for coinciding arguments
in terms of formal series is stated as follows: for every state
\(\Psi \in \VA\) there exists \(N \in \N\) such that
\(\left( z^{\, 2} w^{\, 2} \right)^{N_{\Psi}} V \left( z,w \right) \Psi\)
does not contain negative powers of $z^{\, 2}$ and $z^{\, 2}$
(see \cite{N03} Definition~2.2).
It follows then that $V(z,z)$ is a correctly defined series in $z$ of type
(\ref{eq1.2}) (containing, in general, negative powers of
$z^{\, 2}$) by setting
\(V(z,z) \, \Psi\) $:=$ $\left( z^{\, 2} \right)^{-2N}$
\(\left\{\raisebox{10pt}{\hspace{-3pt}}\right.
\left[\raisebox{10pt}{\hspace{-1pt}}\right.
\left( z^{\, 2} w^{\, 2} \right)^{N}\)
$V \left( z,w \right)$
\(\Psi \left.\raisebox{10pt}{\hspace{-1pt}}\right]
\vrestr{10pt}{z \, = \, w} \left.\raisebox{10pt}{\hspace{-1pt}}\right\}\).
\end{mremark}

The fields $V_{\kappa} \left( z_1,z_2 \right)$ are determined by their correlation
functions. Note that
\begin{equation}
\label{eq3.6}
\lvac V_{\kappa} (z_1 , z_2) \rvac = 0 = \lvac
V_{\kappa} (z_1 , z_2) \, V_{\lambda} (z_3 , z_4) \rvac
\quad \text{for} \quad 0
< \kappa \ne \lambda \,
\end{equation}
since fields of different twists are mutually orthogonal under
vacuum expectation values.
For equal dimensions we can write
\begin{eqnarray}
\label{eq3.8}
&& \lvac V_{\kappa} (z_1 , z_2) \, V_{\kappa} (z_3 , z_4) \rvac \, = \,
(\rho_{13} \rho_{24})^{-\kappa} f_{\kappa} (s,t)
\nonumber \\
&& ( \hspace{2pt} s_{12} \, f_{\kappa} (s,t) \, = \,
t^{-\kappa} f_{\kappa} \left( \frac{s}{t} , \frac{1}{t} \right) \, = \,
f_{\kappa} (s,t) \hspace{2pt} ) \,
\end{eqnarray}
thus obtaining the expansion of the 4-point function
\(\langle 1234 \rangle - \langle 12 \rangle \langle 34 \rangle\)
(see~(\ref{eq2.7})):
\beq\label{exp}
t^{-3} \,
P_4 \left( s,t \right) + B^2 \, s^4
\left(\raisebox{12pt}{\hspace{-2pt}}\right.
1 + \frac{1}{t^4}
\left.\raisebox{12pt}{\hspace{-2pt}}\right)
\, = \,
\mathop{\sum}\limits_{\kappa \, = \, 1}^{\infty} s^{\kappa-1} \,
f_{\kappa} \left( s,t \right)
\, \
\eeq
$B$ being the 2-point normalization constant in~(\ref{eq2.8}).
Moreover, as it is shown in~\cite{DO01} and~\cite{5},
the OPE~(\ref{eq3.2}) implies that $f_{\kappa} \left( s,t \right)$
should have the form:
\begin{eqnarray}
\label{eq3.13}
f_{\kappa} (s,t) = \podr
\frac{1}{u-v} \, \left\{\raisebox{12pt}{\hspace{-3pt}}\right.
F (\kappa - 1, \, \kappa - 1 ; \, 2\kappa - 2 ; \, v) \ g_{\kappa} (u)
\nonumber \\ \podr
- \, F (\kappa - 1, \, \kappa - 1 ; \, 2\kappa - 2 ; \, u) \ g_{\kappa} (v)
\left.\raisebox{12pt}{\hspace{-3pt}}\right\}
\, ,
\end{eqnarray}
\(F(\alpha , \beta ; \gamma ; x) =
\mathop{\sum}\limits_{n \, = \, 0}^{\infty} \,
\txfrac{(\alpha)_n (\beta)_n}{
n! (\gamma)_n} \, x^n = 1 + \txfrac{\alpha\beta}{
\gamma} \, x + \dots\) being Gauss' hypergeometric function
and \(F(0,0;0;u) \equiv 1\); $u$ and $v$ are the ``chiral variables''
of \cite{DO01} (see also~\cite{EPSS}):
\beq\label{st}
s \, = \, uv \, , \quad t \, = \, (1-u)(1-v) \, .
\eeq
The functions $g_{\kappa} \left( u \right)$
should satisfy
\beq\label{g0}
g_{\kappa} \left( 0 \right) \, = \, 0
\eeq
and they determine the
$O_{2\kappa,\ell}$ contributions in~(\ref{eq3.2}) via the expansion
\begin{equation}
\label{eq3.14}
g_{\kappa} (u) = u \, f_{\kappa} (0,1-u) = u \sum_{\ell = 0}^{\infty}
B_{\kappa\ell} \, u^{2\ell} \, F ( 2\ell + \kappa , 2 \ell + \kappa ; 4 \ell + 2
\kappa ; u)
\end{equation}
(the first equality follows from Eqs.~(\ref{g0}), (\ref{eq3.13}) and~(\ref{st})).
The {\it structure
constants} $B_{\kappa \ell}$, unlike those appearing in (\ref{eq3.2}) and
(\ref{eq3.4}), are invariant under rescaling of $O_{2\kappa2\ell}$:
\beq\label{eq3.7}
B_{\kappa \ell} \, := \,
A_{\kappa 2 \ell} \, C_{\kappa 2 \ell}
\quad
( \hspace{1pt} O_{2 \kappa 2 \ell}
\mapsto \lambda \, O_{2 \kappa 2 \ell}
\ \, \Rightarrow \ \,
A_{\kappa} \mapsto \lambda \, A_{\kappa 2 \ell}
, \ \, C_{\kappa 2 \ell} \mapsto \frac{1}{\lambda} \, C_{\kappa 2 \ell}
\hspace{1pt} )
\, .
\eeq
Eqs.~(\ref{exp}) and~(\ref{eq3.13}) allow one to find recurrently the
functions $f_{\kappa} \left( s,t \right)$ as follows:
\beq
\label{eq3.15}
f_1 (0,t) = t^{-3} P_4 (0,t)
, \ \
f_{\kappa} (0,t) = \underset{s \, \to \, 0}{\lim}
\left\{\raisebox{12pt}{\hspace{-3pt}}\right.
s^{1-\kappa} \left[\raisebox{12pt}{\hspace{-2pt}}\right.
t^{-3} P_4 (s,t) - \sum_{\nu = 1}^{\kappa - 1} s^{\nu - 1} f_{\nu} (s,t)
\left.\raisebox{12pt}{\hspace{-2pt}}\right]
\left.\raisebox{12pt}{\hspace{-3pt}}\right\}
,
\eeq
for \(\kappa = 2,3\),
$f_{\kappa} \left( s,t \right)$ being obtained from
the first equation (\ref{eq3.14})
and from Eq.~(\ref{eq3.13}).
For \(\kappa > 3\) second equation (\ref{eq3.15})
should be replaced by the left hand side of Eq.~(\ref{exp}).
It is important to realize that the kernels $K_{\kappa\ell}$
and the hypergeometric functions defining the conformal partial waves (\ref{eq3.14})
are {\it universal}; only the structure constants $B_{\kappa\ell}$ depend on
$\Lagr (z)$ and are, in fact, determined by its 4-point function.

Note that the above algorithm always gives rational solution for
$f_1 \left( s,t \right)$.
Indeed,
by Eq.~(\ref{eq3.13})
\begin{equation}
\label{eq3.12}
f_1 (s,t) = \frac{g(u) - g(v)}{u-v}
\end{equation}
(for \(g \left( t \right) = \left( 1-t \right)^{-3} P_4 \left( 0,1-t \right)\))
is a rational \textit{symmetric} function in $u$ and $v$ which are,
on the other hand, the roots of the second order polynomial equation
\beq\label{uv_eq}
u^2 + \left( t-s-1 \right) u + s \, = \, 0 \quad
(v^2 + \left( t-s-1 \right) v + s \, = \, 0)
\eeq
implied by~(\ref{st}).
This motivates us to consider the field $V_1 \left( z,w \right)$ as
a (\textit{strongly}) \textit{bilocal} field.
For higher twist contributions $V_{\kappa} \left( z,w \right)$,
the function $f_{\kappa} \left( s,t \right)$ is not rational; for instance,
\beq\label{f2}
f_2 \left( s,t \right) \, = \,
\frac{
\ln \left( 1-u \right) g \left( v \right) v
\, - \,
\ln \left( 1-v \right) g \left( u \right) u}{u \,v \left( v-u \right)}
\eeq
and this expression contains a log term for any nonzero function
$g \left( u \right)$.

\msubsection{Symmetrized contribution of twist 2 (conserved)
tensors}{ssec5.2}

The general conformal invariant 3--point function
\(\lvac V_1 (z_1 , z_2) \, O_{2 , \ell} (z_3 ; w) \rvac\)
(\ref{eq3.4})
is harmonic in both $z_1$ and $z_2$ for all $\ell$ which combined
with Wightman positivity and locality
(in particular, the Reeh-Schilder theorem)
implies that:
\begin{equation}
\label{eq3.10}
\Delta_1 \, V_1 (z_1 , z_2) \, = \, 0 \, = \,
\Delta_2 \, V_1 (z_1 , z_2) \quad
( \,
\Delta_j \, = \, \sum_{\mu \, = \, 1}^4 \,
\frac{\partial^2}{(\partial z_j^{\mu})^2} \, )
\, .
\end{equation}
Apart from the assumptions of Wightman positivity and locality
it is shown in \cite{NST03}
that under the expansion~(\ref{eq3.2}) and symmetry~(\ref{eq3.2a})
the equations~(\ref{eq3.10}) are
equivalent to the conservation laws:
\beq\label{conserv}
\di_z \spr \di_w \, O_{2,\ell} \left( z,w \right) \, = \, 0
\, . \
\eeq

At the level of the 4--point function
\(\lvac V_2 \left( z_1,z_2 \right) V_2 \left( z_3,z_4 \right) \rvac\)
the harmonicity is always satisfied.
Indeed, the Laplace equation
in $z_1$
implies
what may be called the {\it conformal Laplace equation}\footnote{The
operator (\ref{eq3.11}) has appeared in various contexts in \cite{EPSS}
\cite{DO01} and \cite{NST03}.} for $f_1$~(\ref{eq3.8}):
\begin{equation}
\label{eq3.11}
\Delta_{st} f_1 (s,t) = 0 \, , \ \Delta_{st} := s \, \frac{\partial^2}{\partial
s^2} + t \, \frac{\partial^2}{\partial t^2} + (s+t-1) \,
\frac{\partial^2}{\partial s \, \partial t} + 2 \left( \frac{\partial}{\partial s} +
\frac{\partial}{\partial t} \right) \, .
\end{equation}
Its general solution is expressed in terms of $u$, $v$~(\ref{st})
by Eq.~(\ref{eq3.12}) which
is a special case of (\ref{eq3.13}) for \(\kappa =1\) (and \(F(0,0;0;u) \equiv 1\)).

We now proceed to writing down the general rational solution
$f_1 \left( s,t \right)$ for our \(d=4\) model.
Accordingly to the first equation (\ref{eq3.15}) and Eq.~(\ref{eq2.7})
it follows that
\beq\label{p1}
p \left( t \right) \, := \, t^3 f_1 \left( 0,t \right)
\eeq
is a polynomial of degree not exceeding 5 and
then the symmetry condition (\ref{eq3.8}) implies:
\beq\label{p}
t^5 \, p \left( \frac{1}{t} \right) \, = \, p \left( t \right)
\quad \Rightarrow \quad
p \left( t \right) \, = \,
\alpha_0 (1+t^5) + \alpha_1 \left( t+t^4 \right) +
\alpha_2 (t^2+t^3)
\, . \
\eeq
Thus the general form of $f_1 \left( s,t \right)$ is a linear combination
of three basic functions constructed by the basic polynomials
\(p_{\nu} \left( t \right) = t^{\nu} + t^{5-\nu}\) for \(\nu = 0,1,2\)
and the algorithm of the previous subsection.

There is a more convenient basis of functions
\(f_1 \left( s,t \right) = j_{\nu} \left( s,t \right)\), again indexed by
\(\nu =0,1,2\), satisfying the requirement that the
$\Ss_3$--symmetrizations of $t^3 \, j_{\nu}$ are ``\textit{eigenfunctions}''
for the following equations:
\beq\label{eigen}
\lambda_{\nu} \left( 1+s_{23}+s_{13} \right)
\left[ \raisebox{9pt}{\hspace{1pt}} t^3 j_{\nu} \left( s,t \right) \hspace{1pt} \right]
\, - \,
t^3 j_{\nu} \left( s,t \right)
\, = \, s^{\sigma_{\nu}} q_{\nu} \left( s,t \right)
\eeq
for \(\sigma_{\nu} \geqslant 1\) where
$q_{\nu} \left( s,t \right)$ are polynomials (of overall
degree \(5-\sigma_{\nu}\)) such that
\(q_{\nu} \left( 0,t \right) \neq 0\), and $s_{23}$, $s_{13}$ are
the $\Ss_3$ generators~(\ref{eq2.9}).
The solutions are given by
\begin{eqnarray}
\label{eq3.18}
j_0 (s,t) \, = \podr j_0 (0,t) = 1+t^{-1}
\, , \nonumber \\
j_1 (s,t) \, = \podr
\left( \frac{1-t}{t} \right)^2 (1+t-s) - 2 \frac{s}{t} \, = \,
j_1 (0,t) - s(1+t^{-2})
\, , \nonumber \\
j_2 (s,t) \, = \podr (1+t^{-3}) [(1+s-t)^2 - s] - 3s (1-t) \, t^{-3} \,
\end{eqnarray}
with ``eigenvalues''  \(\lambda_0 = \lambda_1 = 1\) and
\(\lambda_2 = \txfrac{1}{2}\), and
\(\sigma_0=2\), \(\sigma_1=1\), \(\sigma_2=3\).
The $J_{\nu}$ polynomials~(\ref{eq2.13})--(\ref{eq2.15}) are related to
$j_{\nu}$ as
\beq\label{eq3.19c}
J_{\nu} \left( s,t \right) \, = \,
\lambda_{\nu} \left( 1+s_{23}+s_{13} \right)
\left[ \raisebox{9pt}{\hspace{1pt}} t^3 j_{\nu} \left( s,t \right) \hspace{1pt} \right]
\, . \
\eeq
Thus the equalities~(\ref{eigen}) mean that a 4--point function~(\ref{eq2.7})
obtained by \(P_4 \left( s,t \right) = J_{\nu} \left( s,t \right)\),
for \(\nu =0,1,2\),
corresponds to OPE of
$\Lagr \left( z_1 \right) \Lagr \left( z_2 \right)$
containing twist 2 contribution determined by
\(f_1 \left( s,t \right) = j_{\nu} \left( s,t \right)\) and the
higher twist contributions start with~$\sigma_{\nu}$.

Putting everything together we can, in principle, determine all structure constants
$B_{\kappa\ell}$. It follows from (\ref{eq3.4}) (\ref{eq3.7}) and from the
relation \(A_{\kappa\ell} = N_{\kappa\ell} \, C_{\kappa\ell}\) where
\(N_{\kappa\ell} (>0)\) stands for the normalization of the 2-point function of
$O_{2\kappa\ell}$ that \(B_{\kappa\ell} = N_{\kappa 2\ell} \, C_{\kappa 2\ell}^2\)
should be positive if Hilbert space (or Wightman) positivity holds.
(The full argument uses the classification \cite{M77} of unitary positive energy
representations of $SU(2,2)$ according to which the state spaces spanned by
$O_{2\kappa 2\ell} (z,w) \rvac$, for $\kappa = 1,2,\ldots$, $\ell \geqslant 0$,
belong to the unitary series.) Thus, such a calculation will restrict the
admissible values of the parameters $a_{\nu}$, $b$, $c$ and $B$ in (\ref{eq2.12})
and (\ref{eq3.1}), providing a non-trivial positivity check for the 4-point
function of $\Lagr$. We shall display the corresponding equations and their
solution for \(\kappa =1,2,3\) (the twists for which the 2-point normalization
$B$ does not contribute).

Inserting in the left hand side of (\ref{eq3.14}) for \(\kappa = 1\) the expression
\(f(0,1-u) = \underset{\nu = 0}{\overset{2}{\sum}} a_{\nu} \, j_{\nu} (0,1-u)\)
we can solve with respect to $B_{1\ell}$ with the result
\beq\label{eq3.21}
B_{1\ell} \, = \,
\begin{pmatrix} 4\ell \\ 2\ell \end{pmatrix}^{-1}
\left\{\raisebox{12pt}{\hspace{-3pt}}\right.
2a_0 + 2\ell (2\ell + 1) [2a_1 + (2\ell -1) (\ell + 1) \, a_2 ]
\left.\raisebox{12pt}{\hspace{-2pt}}\right\} \, .
\eeq
The equation for $\kappa = 2$ involves $a_1$, $b$ and $c$:
\begin{eqnarray}
\label{eq3.22}
f_2 (0,1-u) = \podr
a_1 u \left( \frac{1}{(1-u)^3} - 1 \right) + \frac{bu^2}{(1-u)^2}
+ \frac{c}{1-u}
\nonumber \\
= \podr
\sum_{\ell = 0}^{\infty} \, B_{2\ell} \, u^{2\ell}
F (2\ell +2 , 2\ell + 2; 4\ell + 4 ; u) \, .
\end{eqnarray}
Its solution is
\beq\label{eq3.23}
B_{2\ell} \, = \,
\begin{pmatrix} 4\ell + 1 \\ 2\ell \end{pmatrix}^{-1}
\left\{\raisebox{12pt}{\hspace{-3pt}}\right.
\ell (2\ell + 3) [(\ell + 1) (2\ell + 1) \, a_1 + 2b] + c
\left.\raisebox{12pt}{\hspace{-3pt}}\right\} \, .
\eeq

For $\kappa = 3$ we have to use the expression (\ref{eq3.13}) for $f_2 (s,t)$
which involves a log term as
\(F(1,1;2;v) = \underset{n=1}{\overset{\infty}{\sum}}
\txfrac{v^{n-1}}{n} =
\txfrac{1}{v} \, \log \txfrac{1}{1-v}\).
The result is
\begin{eqnarray}
\label{eq3.24}
f_3 (0,1-u) \, = \podr
\left( a_0 + \frac{a_1}{2} \right) (1 + (1-u)^{-3}) - \frac{3}{2}
\, \frac{b}{1-u} \left( 1 + \frac{1}{1-u} \right)
\nonumber \\
\podr + \frac{c}{2} \left\{ \frac{2u-1}{u(1-u)} \left( 1 + \frac{2}{u(1-u)} \right)
- 2 \, \frac{\log (1-u)}{u^3} \right\}
\nonumber \\
= \podr \sum_{\ell = 0}^{\infty} \, B_{3\ell} \, u^{2\ell} \,
F (2\ell + 3 , 2\ell + 3 ; 4 \ell + 6 ; u) \, .
\end{eqnarray}
A computer aided calculation (using Maple) gives in this case
\beq\label{eq3.25}
B_{3\ell} \, = \, \frac{1}{2}
\begin{pmatrix} 4\ell + 3 \\ 2 \ell + 1 \end{pmatrix}^{-1}
\left\{\raisebox{12pt}{\hspace{-2pt}}\right.
(\ell + 1) (2\ell + 3) \, [(\ell + 2) (2\ell + 1) (2a_0 + a_1) - 6b + 4c] - c
\left.\raisebox{12pt}{\hspace{-2pt}}\right\} \, .
\eeq
The positivity of $B_{j\ell}$, $j=1,2,3$ implies
\beq\label{eq3.26}
a_{\nu} \geqslant 0 , \ \, \nu = 0,1,2
\hspace{1pt} ; \ \,
3a_1 + b \geqslant 0
\hspace{1pt} , \ \,
c \geqslant 0
\hspace{1pt} ; \ \,
6 (2a_0 + a_1 - 3b) + 11c \geqslant 0
\hspace{1pt} .
\eeq
This leaves a nonempty domain in the space of (4-point function) parameters in
which positivity holds.

\msubsection{Free field realizations}{ssec5.3n}

Every solution $j_{\nu} (s,t)$ has
a free field realization by a composite field
$V_1^{(\nu)} (z,w)$,
for \(\nu = 0,1,2\), constructed as follows.
Let $\varphi (z)$ and $\psi (z)$ be the
canonical massless scalar and Weyl fields, respectively, introduced in
Sect.~\ref{sec2} and $F_{\mu\nu} (z)$ be the free
electromagnetic (Maxwell) field, characterized by the two point function
\begin{equation}
\label{eq2.18}
\lvac F_{\mu_1 \nu_1} (z_1) \, F_{\mu_2 \nu_2} (z_2) \rvac \, = \,
R_{\mu_1 \mu_2} (z_{12}) \, R_{\nu_1 \nu_2} (z_{12}) - R_{\mu_1 \nu_2} (z_{12})
\, R_{\nu_1 \mu_2} (z_{12})
\end{equation}
where $R_{\mu\nu}$ is related to the vector representation of the conformal
inversion:
\begin{equation}
\label{eq2.19}
R_{\mu\nu} (z) \, = \, \frac{r_{\mu\nu} (z)}{z^{\, 2}}
\, , \quad
r_{\mu\nu} (z) \, = \, \delta_{\mu\nu} - 2 \, \frac{z_{\mu} z_{\nu}}{z^{\, 2}}
\, , \quad
(r^2)_{\mu\nu} \, = \, \delta_{\mu\nu} \, .
\end{equation}
Then the bilocal fields:
\beqa\label{nu0}
\hspace{-40pt}
& V_1^{(0)} \! (z_1 , z_2) & \hspace{-8pt} = \
: \hspace{-1pt} \varphi (z_1) \, \varphi (z_2) \hspace{-2pt} :
\hspace{1pt} ,
\\ \label{eq4.11} \hspace{-40pt}
& V_1^{(1)} \! (z_1 , z_2) & \hspace{-8pt} = \
: \hspace{-1pt} \psi^+ (z_1) \, {\dirz}_{12} \, \psi (z_2) \hspace{-2pt} :
\, - \,
: \hspace{-1pt} \psi^+ (z_2) \, {\dirz}_{12} \, \psi (z_1) \hspace{-2pt} :
\hspace{1pt} ,
\\ \label{nu2} \hspace{-40pt}
& V_1^{(2)} \! (z_1 , z_2) & \hspace{-8pt} = \
\frac{1}{4} \, \rho_{12}
: \hspace{-1pt} F^{\sigma\tau} (z_1) F_{\sigma\tau} (z_2) \hspace{-2pt} :
- \ \delta^{\sigma\tau} \, z_{12}^{\mu} \, z_{12}^{\nu}
: \hspace{-1pt} F_{\sigma\mu} (z_1) F_{\tau\nu} (z_2) \hspace{-2pt} :
\hspace{1pt}
\eeqa
($:\_ :$ being the standard free field's normal product)
or the sum of commuting copies of such expressions,
are harmonic, symmetric under exchange of $z_1$ and $z_2$, and give
an operator realization of the corresponding dimensionless 4-point
functions $j_{\nu} (s,t)$~(\ref{eq3.18}).
Indeed, the cases of \(\nu =0\) and \(\nu = 2\) have been already
proved in \cite{NST02} and \cite{NST03}, and the corresponding calculation
for \(\nu = 1\) is done in Appendix~A.

Moreover, for every \(\nu =0,1,2\) there is a composite field realization
of the scalar field \(\Lagr (z) \equiv \Lagr^{(\nu)} (z)\)
reproducing the OPE~(\ref{eq3.1}) with twist two part presented
by the bilocal field $V_1^{(\nu)} (z,w)$ as well as
the truncated 4-point function corresponding to
$J_{\nu} (s,t)$~(\ref{eq2.13})--(\ref{eq2.15}) up to normalization.

For \(\nu = 2\) it was shown in \cite{NST03} that the free Maxwell Lagrangian
\begin{equation}
\label{eq2.17}
\Lagr_0 (x) \, = \, - \frac{1}{4} : F_{\mu\nu} (z) \, F^{\mu\nu} (z) : \,
\end{equation}
does the job.
(The calculation proving that Eqs.~(\ref{eq2.17}), (\ref{eq2.18}) and (\ref{eq2.19})
yield a $w_4^t$ proportional to
$\rho_{13}^2$ $\rho_{24}^2$ $(\rho_{12}$ $\rho_{23}$ $\rho_{34}$
$\rho_{14})^{-3}$ $J_2 (s,t)$ is given in Appendix B of \cite{NST02}.)

For \(\nu = 0\) and \(\nu = 1\) one has to introduce an additional (independent)
generalized free field\footnote{K.-H. Rehren, private communication.}.
Let $\phi (z)$ be a generalized free scalar neutral field of dimension 3
commuting with $\varphi (z)$. Then the composite scalar field of dimension~4:
\beqa\label{L0}
\Lagr^{(0)} (z) \, := \,
\varphi (z) \, \phi (z) \, ( \, \equiv \,
: \hspace{-1pt} \varphi (z) \, \phi (z) \hspace{-1pt} : \, )
\eeqa
has the OPE:
\beqa\label{OPE0}
\Lagr^{(0)} (z_1) \, \Lagr^{(0)} (z_2) \, = \podr
\frac{B_{\varphi}B_{\phi}}{\rho_{12}^4}
\, + \,
\frac{B_{\phi}}{\rho_{12}^3} \, V_1^{(0)} (z_1 , z_2)
\, + \,
\frac{B_{\varphi}}{\rho_{12}}
: \hspace{-1pt} \phi (z_1) \, \phi (z_2) \hspace{-1pt} :
\nn
\podr + \,
: \hspace{-1pt} \Lagr^{(0)} (z_1) \, \Lagr^{(0)} (z_2) \hspace{-1pt} :
\eeqa
(using the standard Wick normal product).
Similarly,
let $\chi (z)$ be a generalized free Weyl field of dimension
$\txfrac{5}{2}$,
with 2-point function
\beq\label{chi_2pt}
\lvac \chi \left( z_1 \right) \chi^+ \left( z_2 \right) \rvac
\, = \,
\rho_{12}^{-3} \, \dirz_{12}
\eeq
and anticommuting with $\psi (z)$.
Then the composite field:
\beq\label{L1}
\Lagr^{(1)} (z) \, := \,
\psi^+ (z) \, \chi (z) \, + \, \chi^+ (z) \, \psi (z)
\eeq
has the OPE:
\beqa\label{OPE1}
\Lagr^{(1)} (z_1) \, \Lagr^{(1)} (z_2) \, = \podr
\frac{1}{\rho_{12}^4}
\, + \,
\frac{1}{\rho_{12}^3} \, V_1^{(1)} (z_1 , z_2)
\, + \,
\frac{1}{\rho_{12}^2}
\left\{\raisebox{10pt}{\hspace{-3pt}}\right.
: \hspace{-1pt} \chi^+ (z_1) \, {\dirz}_{12}^+ \, \chi (z_2) \hspace{-2pt} :
\nn
\podr - \,
: \hspace{-1pt} \chi^+ (z_2) \, {\dirz}_{12}^+ \, \chi (z_1) \hspace{-2pt} :
\left.\raisebox{10pt}{\hspace{-3pt}}\right\}
\, + \,
: \hspace{-1pt} \Lagr^{(1)} (z_1) \, \Lagr^{(1)} (z_2) \hspace{-1pt} :
\, . \
\eeqa
It follows then that the $V_1^{(\nu)}$ correspond to the twist two parts
in the OPE of $\Lagr^{(\nu)}$ (\(\nu = 0,1\)).
Note that the third terms in the OPE's~(\ref{OPE0}) and (\ref{OPE1}) do not
correspond to $\rho_{12}^{-1} V_3$ and $\rho_{12}^{-2} V_2$, respectively,
since they have rational correlation functions, while the 4-point functions
of $V_{\kappa}$ for \(\kappa \geqslant 2\) involve log terms (cf.~(\ref{f2})).

A straightforward computations show that the 4-point functions of
$\Lagr^{(0)}$ and $\Lagr^{(1)}$ are proportional to
$J_0$ and $J_1$ (resp.).
For $J_1$ an explicit calculation is made in Appendix~A.
We will give also in the next section a general argument
(see Proposition~\ref{prp6.1}) that all the $2n$--point functions
of $\Lagr^{(\nu)}$ for \(\nu = 0,1,2\) are reproduced,
up to a multiplicative constant, by the $2n$-point function
of the corresponding $V_1^{(\nu)}$ using a generalization of the
symmetrization procedure of~(\ref{eigen}).

\msection{Towards constructing nontrivial GCI QFT mo\-dels}{sec6n}

\msubsection{The symmetrization ansatz}{ssec6.1n}

We observe that the twist two contribution to the $2n$-point function,
\begin{eqnarray}
\label{eq3.27}
&& w_1 (1,2 ; 3,4 ; \ldots ; 2n - 1,2n) \, := \,  \nonumber \\
&&
\left(\raisebox{12pt}{\hspace{-2pt}}\right.
\prod\raisebox{-2.5pt}{\(
\raisebox{12.5pt}{\hspace{-1pt}}_{i=1}^{n} \ \rho_{2i-1,2i}^{-3}\)}
\left.\raisebox{12pt}{\hspace{-2pt}}\right)
\lvac V_1 (z_1 , z_2)
\, V_1 (z_3 , z_4) \ldots V_1 (z_{2n-1}, z_{2n}) \rvac \, ,
\qquad
\end{eqnarray}
combined with locality, implies the existence of higher twist terms.
The question arises is there a possibility to generate the full truncated
correlation functions of the model by an appropriate symmetrization of such
a bilocal field contribution.
The difficulty in making this idea precise is that after the
permutation symmetrization of $w_1 (1,2 ; 3,4 ; \ldots ; 2n - 1,2n)$
its twist two part may not be represented by the initial $w_1$.
Indeed, we have already seen in Sect.~\ref{ssec5.2}
that for the dimensionless 4-point function $j_{\nu} \left( s,t \right)$
we should use permutation symmetrizations with different normalization
($\lambda_{\nu}$, see Eq.~(\ref{eigen})) in order to obtain the truncated
(dimensionless) function $P_4$ whose twist two part is again~$j_{\nu}$.
Moreover, for higher point functions it may happen that the
generalized ``eigenvalue'' problem~(\ref{eigen}) does not
have a solution spanning the whole space of possible twist two contributions.
We shall say that $V_1 \left( z,w \right)$ is symmetrizable if its correlation
functions belong to the resulting subspace.

To be more precise, instead of the functions $w_1$~(\ref{eq3.27}) we will use
what may be called \textit{truncated} $2n$-point function of bilocal
fields, setting
\begin{eqnarray}
\label{eq3.28bis}
w_{1}^{t} (1,2; \ldots ;2n-1,2n) \, = \podr w_{1}
(1,2; \ldots ;2n-1,2n) \quad \text{for} \quad  n < 4
\, , \nonumber \\
w_{1}^{t} (1,2; \ldots ;7,8) \, = \podr
w_{1}(1,2; \ldots ;7,8) - w_{1} (1,2;3,4) \, w_{1}
(5,6;7,8) \nonumber \\
\podr - \, w_{1} (1,2;5,6) \, w_{1} (3,4;7,8)
\nonumber \\
\podr - \, w _{1} (1,2;7,8) \, w_{1} (3,4;5,6)
\end{eqnarray}
(and similar expressions involving symmetric subtractions for \(n > 4\)).
We say that $V_1$ is {\it symmetrizable} if for any \(n = 2,3,\ldots\),
there is a $\lambda_n$ such that for all \(i=1,\ldots,n-1\):
\beqa\label{eq3.28}
&
\underset{\rho_{2i-1,2i} \, \to \, 0}{\lim} \,
\left\{\raisebox{10pt}{\hspace{-3pt}}\right.
\rho_{2i-1,2i}^3 (w^t (z_1 , \ldots ,
z_{2n}) - w_1^t (1,2 ; \ldots ; 2n-1,2n))
\left.\raisebox{10pt}{\hspace{-4pt}}\right\} \, = \, 0
\, , \qquad
& \\ \label{eq3.29bis} &
w^t (1,2,\ldots , 2n) \, = \,
\lambda_n \text{\Large $\sum$}'
\, w_1^t (1 , i_2 ; \ldots ; i_{2n-1} , i_{2n})
\, , \
&
\eeqa
where the sum $\text{$\sum$}'$ is spread over all $(2n-1)!!$ permutations
\((1,2,\ldots ,2n) \mapsto (1,i_2,\ldots ,i_{2n})\)
whose entries satisfy the inequalities
\beq\label{ineq}
1 \equiv i_1 < i_2 \, ,\,  \dots \, ,\, i_{2n-1} < i_{2n}
\ \ \text{and} \ \
i_1 < i_3 < \dots < i_{2n-1}
.
\eeq
The odd point truncated functions \(w_{1}^{t} (1,\ldots , 2n-1)\)
are assumed to be zero.
(This assumption will be justified for a gauge field theory
Lagrangian in Sect.~\ref{ssec6.3n} below.) Eqs.~(\ref{eq3.28}) and (\ref{eq3.29bis})
tell us that $w^t(1,2,\dots ,2n)$ involves the same twist two contribution for
any pair of arguments.

It can be demonstrated, using the free fields realizations, that
all bilocal fields $V_1^{(\nu)}$ are symmetrizable.

\begin{mproposition}\label{prp6.1}
The bilocal fields $V_1^{(\nu)} (z,w)$ of the models of
$\Lagr^{(\nu)} (z)$, introduced in Sect.~\ref{ssec5.3n},
are symmetrizable.
\end{mproposition}

\noindent
\textit{Sketch of the proof.}
By the Wick theorem, it is obvious that the odd point correlation functions
of $\Lagr^{(0)} (z)$ and $\Lagr^{(1)} (z)$ vanish.
For $\Lagr^{(2)} (z)$ this is implied by the ``electric-magnetic''
(or Hodge) duality (see for more details Sect.~\ref{ssec6.3n}).
Observing that the fields $\Lagr^{(\nu)} (z)$ and $V_1^{(\nu)} (z_1,z_2)$
have the following general structure:
\beqa\label{gen_str}
\Lagr^{(\nu)} (z) \, = \podr
A_{\nu} \, \mathop{\text{\small $\sum$}}\limits_{a} \,
\left(\raisebox{10pt}{\hspace{-3pt}}\right.
: \hspace{-1pt} \vartheta_a^* (z) \, \sigma_a (z) \hspace{-2pt} :
\, + \,
: \hspace{-1pt} \sigma_a^* (z) \, \vartheta_a (z)  \hspace{-2pt} :
\left.\raisebox{10pt}{\hspace{-3pt}}\right)
,
\nn
V_1^{(\nu)} (z_1,z_2) \, = \podr
B_{\nu} \, \rho_{12}^3 \, \mathop{\text{\small $\sum$}}\limits_{a,\, b} \,
\left(\raisebox{10pt}{\hspace{-3pt}}\right.
\lvac \sigma_a (z_1) \, \sigma_b^* (z_2) \rvac
: \hspace{-1pt} \vartheta_a^* (z_1) \, \vartheta_b (z_2) \hspace{-2pt} :
\nn
\podr + \,
\lvac \sigma_a^* (z_1) \, \sigma_b (z_2) \rvac
: \hspace{-1pt} \vartheta_a (z_1) \, \vartheta_b^* (z_2) \hspace{-2pt} :
\left.\raisebox{10pt}{\hspace{-3pt}}\right)
\nn && \hspace{-78pt} \mgvspc{12pt} \hspace{-9pt}
( \, \text{for} \ \, \nu = 0 : \ \,
\vartheta = \varphi,\ \, \sigma = \phi \, ; \ \,
\text{for} \ \, \nu = 1 : \ \,
\{\vartheta_a\} = \psi,\ \, \{\sigma_a\} = \chi \, ;
\nn && \hspace{-78pt} \hspace{68pt} \hspace{-9pt}
\text{for} \, \ \nu = 2 : \ \,
\{\vartheta_a\} = \{\sigma_a\} = \{F_{\mu\nu}\} \, ) \, ,
\eeqa
where
$A_{\nu}$ and $B_{\nu}$ are constants,
it then follows,
by the Wick theorem, that the truncated $2n$--point function of
$\Lagr^{(\nu)} (z)$ is a sum of $1$--loop contributions which is
proportional to the symmetrization of the type of~(\ref{eq3.29bis}) of the
$2n$--point function $w_1^t$~(\ref{eq3.28bis}).$\quad\Box$

\begin{mremark}\label{rm6.1}
We note that the electro-magnetic Lagrangian $\Lagr^{(2)}$
has an additional symmetry under the exchange of the fields $F_{\mu\nu}$
in comparison with $\Lagr^{(0)}$ and $\Lagr^{(1)}$.
This leads to the fact that the eigenvalues $\lambda_n$ for
$\Lagr^{(0)}$ and $\Lagr^{(1)}$ are equal and
twice bigger than those for $\Lagr^{(2)}$.
\end{mremark}

\msubsection{Elementary contributions to the truncated
$2n$-point functions}{ssec6.2n}

The $2n$-point function of the composite field $V_1^{(1)}$ has a simple
structure, verified for \(n=2,3,4\) (see Appendix)
and conjectured
for all $n$.

We begin by illustrating this structure for \(n=2\) and $3$.
First note that the general $2n$-point function of a bilocal field
$V_1 (z_1,z_2)$ should have an $2^n n!$--element symmetry group of
 permutations of the arguments. This group is the
\((\Z_2)^{\times n} \times \Ss_n\) subgroup
of $\Ss_{2n}$ consisting of exchanging the arguments of each individual
$V_1 (z_{2k-1},z_{2k})$ and of permuting the $V$'s.

The 4-point function of $V_1^{\left( 1 \right)}$ (presented by
$j_1(s,t)$) is a sum of two terms having different \textit{pole structure}.
These are
\beq\label{e6.7n}
{\mathcal W} (12;34) \, = \,
2 \, \frac{\rho_{13} \rho_{24} - \rho_{12} \rho_{34} - \rho_{14} \rho_{23}}{
\rho_{14}^2 \rho_{23}^2} \,
\eeq
and its permutation of the arguments ${\mathcal W} (12;43)$.
We will call \textit{elementary} such contributions.
For the 6-point function we have 8 different pole structures,
i.e. elementary contributions,
forming a single orbit under the action of the group
\((\Z_2)^{\times 3} \times \Ss_3\).
One of these structures is:
\begin{eqnarray}
\label{e6.8n}
&& \hspace{-20pt}
{\mathcal W} (12 ; 34 ; 56) = (\rho_{16}
\rho_{23} \rho_{45})^{-2} \bigl\{ \rho_{12} (\rho_{34} \rho_{56} - \rho_{35}
\rho_{46} + \rho_{36} \rho_{45})
\nonumber \\
&& \hspace{-20pt}
- \, \rho_{13} (\rho_{24} \rho_{56} - \rho_{25} \rho_{46} + \rho_{26} \rho_{45}) +
\rho_{14} (\rho_{23} \rho_{56} - \rho_{25} \rho_{36} + \rho_{26} \rho_{35})
\nonumber \\
&& \hspace{-20pt}
- \, \rho_{15} (\rho_{23} \rho_{46} - \rho_{24} \rho_{36} + \rho_{26} \rho_{34}) +
\rho_{16} (\rho_{23} \rho_{45} - \rho_{24} \rho_{35} + \rho_{25} \rho_{34}) \bigl\}
\,
\end{eqnarray}
and it has a \(\Z_3 \times \Z_2\) symmetry generated
by the cyclic permutation $(1,$ $2,$ $3,$ $4,$ $5,$ $6)$
$\mapsto$ $(3,$ $4,$ $5,$ $6,$ $1,$ $2)$ and the
inversion $(1,$ $2,$ $3,$ $4,$ $5,$ $6)$ $\mapsto$ $(6,$ $5,$ $4,$ $3,$ $2,$ $1)$.
Thus the number of the orbit's elements is indeed
\(8 = \txfrac{2^3 \times 3!}{2 \times 3} = 2^2 2!\).

Now we observe that the numerators of the expressions~(\ref{e6.7n}) and (\ref{e6.8n})
have a \textit{Wick structure} of fermionic correlation functions with
a ``propagator'' \([12] := \rho_{12} \equiv [21]\).
This observation is confirmed also for the 8-point function
of $V_1^{(1)}$ and we conjecture it for all $2n$.

Multiplying each elementary contribution to the $2n$--point
truncated correlation function of $V_1^{(1)}$ by the prefactor of~(\ref{eq3.27})
we obtain an elementary contribution
to the correlation function of $\Lagr^{(1)}$.
For instance, the 6--point truncated function of $\Lagr^{(1)}$
we have 120 elementary contributions organized in 15 sums of
8 element contributions of type $w_1^t$~(\ref{eq3.27}).


\refstepcounter{subsection}
\subsection*{\bf\normalsize \arabic{section}.\arabic{subsection}.
	Is there a non-trivial gauge field theory model?
	Restrictions on the parameters in the 4-point function}\label{ssec6.3n}%
\addcontentsline{toc}{subsection}{\arabic{section}.\arabic{subsection}.
	Is there a non-trivial gauge field theory model?}

We now address the question how to characterize the local gauge invariant
Lagrangian, which gives rise to a 4-form
\begin{equation}
\label{eq4.1}
\Lagr (z) \, dz_1 \wedge dz_2 \wedge dz_3 \wedge dz_4 = \TRACE \, (^*F(z)
\wedge F(z)) \, ,
\end{equation}
where $F$ is the (Maxwell, Yang-Mills) curvature 2-form and $^*F$ is its Hodge
dual:
\begin{equation}
\label{eq4.2}
F(z) = \frac{1}{2} \, F_{\mu\nu} (z) \, dz^{\mu} \wedge dz^{\nu} \, , \ ^*F (z) =
\frac{1}{4} \, \varepsilon_{\kappa \lambda \mu \nu} \, F^{\kappa \lambda} \,
dz^{\mu} \wedge dz^{\nu} \, ,
\end{equation}
without introducing gauge dependent quantities like $F$ (in the non-abelian case)
or the connection 1-form
(of the gauge potential) $A$. We first note that a pure gauge Lagrangian of type
(\ref{eq4.1}) ({\it i.e.} a Lagrangian without matter fields) should not allow for
a scalar of dimension ($=$ twist) 2 in the OPE of
$\Lagr (z_1) \, \Lagr (z_2)$. In view of (\ref{eq3.21}) this implies
\begin{equation}
\label{eq4.3}
a_0 = 0 \, .
\end{equation}
Furthermore, assuming invariance of the theory under ``electric-magnetic'' (or
Hodge) duality\footnote{We thank Dirk Kreimer  for a discussion on this point.},
and noting that $^* (^*F) = -F$ in Minkowski space, we deduce that the theory
should be invariant under a change of sign of $\Lagr$. Hence all odd-point
functions of $\Lagr$ should vanish. We make the stronger assumption that {\it no} scalar field of dimension 4 should
appear in the OPE of two $\Lagr$'s. According to (\ref{eq3.23}) this implies
\begin{equation}
\label{eq4.4}
c = 0 \, .
\end{equation}
(The vanishing of the 3-point function of the Maxwell Lagrangian is verified by a
direct calculation.) We are thus left with the 3 parameters $a_1$, $a_2$, and $b$
in the truncated 4-point function, the positivity restrictions (\ref{eq3.26})
implying
\begin{equation}
\label{eq4.5}
a_1 \geqslant 0 \, , \quad
a_2 \geqslant 0 \, , \quad
a_1 + a_2 > 0 \, , \quad -3a_1 \leqslant b \leqslant \frac{1}{3} \, a_1 \, .
\end{equation}
Clearly for $a_1 = 0$ we shall also have $b=0$ and the truncated 4-point function
will be a multiple of that of the free electromagnetic Lagrangian (\ref{eq2.17})
(Sect.~\ref{ssec5.3n}). In order to go beyond the free field theory we shall assume
$a_1 > 0$.

It appears that any ``{\it minimal model}''~--~generated by the bilocal
fields $V_1$ according to the symmetrization ansatz of
Sect.~\ref{ssec6.1n}~--~corresponds
to an $\Lagr$ that is a sum of normal products of generalized
free fields. (It would be interesting to give a proof of this conjecture.)
A careful analysis shows that there is no free field realization of the 4-point
function with a nonvanishing $b$ (but \(a_0=0=c\)). It is the resulting
3-parameter family of models that is most attractive from our point
of view and deserves a systematic study.

\msection{Concluding remarks}{sec7}

Global conformal invariance \cite{NT01} opens the way of constructing 4- (or
higher) dimensional QFT models satisfying all Wightman axioms (except for
asymptotic completeness). It also allows to construct elliptic correlation functions for finite temperature equilibrium states and to display modular properties of energy mean values.

Experience with gauge field theory suggests that the
simplest local gauge invariant observable is the Lagrangian density $\Lagr$.
The present update of our effort to construct a non-perturbative GCI gauge QFT
\cite{NST02} \cite{NST03} displays some new features and suggests new questions (or
new ways of approaching old ones).

-- We emphasize that the main tool for attacking the difficult problem of Wightman
positivity are the {\it conformal partial wave expansions} of 4-point functions.
They should be extended to 4-point functions of composite (tensor) fields or,
alternatively, to higher point functions of $\Lagr (z)$. OPE provide just a
means to derive such expansions with invariant under rescaling structure constants
(like (\ref{eq3.7})).

-- The notion of a symmetrizable strongly bilocal field $V_1(z_1,z_2)$,
which is harmonic in each argument, is introduced (in Sect.~\ref{sec6n}).

-- It is demonstrated that all twist two contributions to the 4-point
function can be realized as normal products of free fields.

It seems possible~--~and it would be worthwhile the effort of
proving~--~that $\Lagr(z)$ for a ``minimal model'' would
be itself a sum of normal products of free fields.
In this case one should concentrate on
studying a theory with truncated
4--point function of $\Lagr$ given by (\ref{eq2.7}) (\ref{eq2.12})
with \(a_0=0=c\) but with a non-zero~$b$.
Such a conformal model still has a chance to describe a gauge field Lagrangian
that is not a part of a free field theory.

\bigskip

\noindent {\bf Acknowledgments.}
We have benefited by discussions with Yassen Stanev together with whom most of
the results of Sects. \ref{sec4}-\ref{sec6n} were obtained (\cite{NST03}),
and with Karl-Henning Rehren whose contribution to our ongoing joint work on this
topic is reflected in Sect.~\ref{ssec5.3n}.
I.T. acknowledges the hospitality of l'Institut des Hautes Etudes Scientifiques
(Bures-sur-Yvette) where a first draft of this paper was written and the support
of the Alexander von Humboldt Foundation and the hospitality of the Institut
f{\"u}r Theoretische Physik der Universit{\"a}t G{\"o}ttingen while its final
version was completed.
Both authors acknowledge partial support by the Research Training Network within
Framework Programme 5 of the European Commission under contract HPRN-CT-2002-00325.


\appendix
\apsection{Computing correlation functions of $V_1^{(1)}(z_1 , z_2)$}{ap1}

The $4$-point function $j_1 (s,t)$ for $V_1^{(1)}$ given by (\ref{eq4.11}) and the
$2$-point function of $\psi$ normalized according to (\ref{w_2pt}) is expressed
as a symmetric combination of traces:
\begin{eqnarray}
\label{eqA.1}
\lvac V_1^{(1)} (z_1 , z_2) \, V_1^{(1)} (z_3 , z_4) \rvac = \podr
\frac{\TRACE \{ \dirz_{12} (\dirz_{23}^+ \dirz_{34} \dirz_{14}^+ + \dirz_{14}^+ \dirz_{34} \dirz_{23}^+)\}}{\rho_{14}^2 \, \rho_{23}^2} \nonumber \\
\podr - \,
\frac{\TRACE \{ \dirz_{12} (\dirz_{24}^+ \dirz_{34}
\dirz_{13}^+ + \dirz_{13}^+ \dirz_{34} \dirz_{24}^+)\}}{
\rho_{13}^2 \, \rho_{24}^2} \, .
\end{eqnarray}
It is sufficient to compute the first term since the second can be obtained from it
by the substitution \(z_3 \rightleftarrows z_4\). To do that we shall use the
following trace formula for the product of any four 4-vectors $a$, $b$, $c$, $d$
written as quaternions:
\begin{eqnarray}
\label{eqA.2}
& \TRACE (\dira \, \dirb^+ \, \dirc \, \dird^+)
\, = \, 2 [(ab)(cd) - (ac)(bd) + (ad)(bc) + \det (a,b,c,d)] \, ,
& \nonumber \\ &
2(ab) \, = \, \TRACE \, \dira \, \dirb^+ \, , &
\end{eqnarray}
$\det (a,b,c,d)$, the determinant of the $4 \times 4$ matrix of the components
of the four (column) vectors, changing sign under transposition of any two arguments.
It follows that
\begin{eqnarray}
\label{eqA.3}
& \TRACE (\dirz_{12} \dirz_{23}^+ \dirz_{34} \dirz_{14}^+ +
\dirz_{12} \dirz_{14}^+ \dirz_{34} \dirz_{23}^+)
\nonumber \\
& = \, 4 [(z_{12} z_{23}) (z_{34} z_{14}) - (z_{12} z_{34}) (z_{14} z_{23}) + (z_{12} z_{14}) (z_{23} z_{34})] \, .
\end{eqnarray}
To reproduce (\ref{e6.7n}) one uses the relations
\beqa & \hspace{-20pt}
2 \, z_{12} z_{23} =
\rho_{13} - \rho_{12} - \rho_{23} \, , \ 2 \, z_{34} z_{14} =
\rho_{34} + \rho_{14} - \rho_{13}
, \ \text{etc.}
\nonumber \\ \label{eqA.4} & \hspace{-20pt}
2 \, z_{12} z_{34} =
\rho_{14} + \rho_{23} - \rho_{13} - \rho_{24} \, , \ 2 \, z_{14} z_{23} =
\rho_{13} + \rho_{24} - \rho_{12} - \rho_{34} , \ \text{etc.}
\eeqa
Similarly, the (polynomial) {\it elementary contribution} (\ref{e6.8n})
to the 6-point function
$$
\rho_{16}^2 \rho_{23}^2 \rho_{45}^2 \,
\lvac V_1^{(1)} (z_1 , z_2) \, V_1^{(1)} (z_3 , z_4) \, V_1^{(1)} (z_5 , z_6) \rvac
$$
is given by
\begin{eqnarray}
\label{eqA.5}
&
P (12; 34; 56) := \TRACE \{ \dirz_{12} (\dirz_{23}^+ \dirz_{34} \dirz_{45}^+ \dirz_{56} \dirz_{16}^+ + \dirz_{16}^+ \dirz_{56} \dirz_{45}^+ \dirz_{34} \dirz_{23}^+)\}
& \nonumber \\ &
= 4 \{ (z_{12} z_{23}) [ (z_{34} z_{45}) (z_{56} z_{16}) - (z_{34} z_{56}) (z_{45} z_{16}) + (z_{34} z_{16}) (z_{45} z_{56})]
& \nonumber \\ &
- (z_{12} z_{34}) [ (z_{23} z_{45}) (z_{56} z_{16}) - (z_{23} z_{56}) (z_{45} z_{16}) + (z_{23} z_{16}) (z_{45} z_{56})] + \ldots
& \qquad  \nonumber \\ &
+ (z_{12} z_{16}) [ (z_{23} z_{39}) (z_{45} z_{56}) - (z_{23} z_{45}) (z_{34} z_{56}) + (z_{23} z_{56}) (z_{34} z_{45})] \}
&
\end{eqnarray}
($5 \times 3$ terms). Applying to this expression the relations of type
(\ref{eqA.4}) (and using Maple to simplify the result) we recover (\ref{e6.8n}).


\end{document}